\newif\ifconfver
\confverfalse      

\newif\ifcutshort      
\cutshorttrue

\newif\ifcutshortlvltwo  
\cutshortlvltwofalse

\ifconfver
\documentclass[10pt,twocolumn,twoside]{IEEEtran}
\else
\documentclass[11pt, draftclsnofoot, onecolumn]{IEEEtran}
\fi

\usepackage{amsmath,amsfonts,bm,amssymb,epsfig,psfrag,ifthen,color}
\usepackage{booktabs}
\usepackage{multirow}
\usepackage[table]{xcolor}
\usepackage{array}
\usepackage{cite}
\usepackage{calc}
\usepackage{caption}
\usepackage{graphicx}
\usepackage{psfrag}
\usepackage[tight,footnotesize]{subfigure}
\usepackage{stfloats}
\usepackage{url}
\usepackage{longtable}
\usepackage{amsmath,amsfonts,amssymb,amsbsy,bm,paralist,theorem,ifthen,color}
\usepackage{algorithmic,algorithm,theorem}
\usepackage{mathtools}
\usepackage{threeparttable}
\usepackage{diagbox}
\usepackage{wrapfig}
\usepackage{hyperref}

\newcommand\Dc{\ensuremath{\mathcal{D}}}

\newcommand\Sc{\ensuremath{\mathcal{S}}}

\newcommand\Bc{\ensuremath{{\mathcal{B}}}}
\newcommand\Cc{\ensuremath{{\mathcal{C}}}}

\newcommand\Jc{\ensuremath{{\mathcal{J}}}}

\newcommand\Oc{\ensuremath{{\mathcal{O}}}}

\newcommand\xb{\ensuremath{{\bf x}}}

\newcommand\yb{\ensuremath{{\bf y}}}

\newcommand\Hb{\ensuremath{{\bf H}}}

\newcommand\Ab{\ensuremath{{\bf A}}}

\newcommand\ob{\ensuremath{{\bf o}}}

\newcommand\Xb{\ensuremath{{\bf X}}}
\newcommand\Yb{\ensuremath{{\bf Y}}}
\newcommand\Ub{\ensuremath{{\bf U}}}

\newcommand\Wb{\ensuremath{{\bf W}}}

\newcommand\xib{\ensuremath{{\bm \xi}}}

\newcommand\E{\ensuremath{{\mathbb{E}}}}

\newcommand\oneb{\ensuremath{{\bf 1}}}

\newcommand\tr{\ensuremath{{\rm Tr}}}

\newcommand\Rbb{\ensuremath{{\mathbb{R}}}}

\newcommand{\wt}{\widetilde}

\newcommand{\ol}{\overline}

\newtheorem{Lemma}{Lemma}

\newtheorem{Theorem}{Theorem}
\newtheorem{Def}{Definition}

\newtheorem{assumption}{Assumption}
\newtheorem{Remark}{Remark}

\ifconfver

\else

\fi

\newcommand{\tabincell}[2]{\begin{tabular}{@{}#1@{}}#2\end{tabular}}

\hypersetup{colorlinks=true,linkcolor=black,citecolor=black,urlcolor=black}

\allowdisplaybreaks[4]
\begin{document}

\bibliographystyle{IEEEtran}

\title{ {Differentially Private Federated Clustering over Non-IID Data}}

\ifconfver \else {\linespread{1.1} \rm \fi

\author{Yiwei~Li,   Shuai~Wang,   Chong-Yung~Chi,~\IEEEmembership{Life Fellow,~IEEE,}
Tony Q. S.~Quek,~\IEEEmembership{Fellow,~IEEE}
\thanks{This work   is supported by the Ministry of Science and Technology, Taiwan,
under   Grants MOST 111-2221-E-007-035-MY2 and MOST 110-2221-E-007-031.  It is also supported in part by the National Research Foundation, Singapore and Infocomm Media Development Authority under its Future Communications Research \& Development Programme.   }
\IEEEcompsocitemizethanks{\IEEEcompsocthanksitem Y. ~Li and C.-Y.~Chi are with Institute of Communications Engineering, National Tsing Hua University, Hsinchu 30013, Taiwan (e-mail: lywei0306@foxmail.com,~cychi@ee.nthu.edu.tw).\protect
\IEEEcompsocthanksitem S. ~Wang is  with Information Systems Technology and Design,
Singapore University of Technology and Design,  Singapore 487372 (e-mail: shuaiwang@link.cuhk.edu.cn).
\protect
\IEEEcompsocthanksitem T. Q. S.~Quek is  with the Singapore University of Technology and Design, Singapore 487372, and also with the Yonsei Frontier Lab, Yonsei University, South Korea (e-mail: tonyquek@sutd.edu.sg).    }}

\maketitle
\begin{abstract}
In this paper, we investigate federated clustering (FedC) problem,  that aims to accurately partition unlabeled data samples distributed over massive clients into finite clusters  under the orchestration of a parameter server, meanwhile considering data privacy.
Though it is an NP-hard optimization problem involving real variables denoting cluster centroids and binary variables denoting the cluster membership of each data sample, we judiciously reformulate the FedC problem into a  non-convex optimization problem with only one convex constraint, accordingly yielding a soft clustering solution. Then a novel FedC algorithm  using differential privacy (DP) technique, referred to as DP-FedC, is proposed in which partial clients participation and multiple local model updating steps are also considered. Furthermore, various attributes of the proposed DP-FedC are obtained through theoretical analyses of privacy protection and convergence rate, especially for the case of non-identically and independently distributed (non-i.i.d.) data, that ideally serve as the guidelines for the design of the proposed DP-FedC. Then some experimental results on two real datasets are provided to demonstrate the efficacy of the proposed  DP-FedC together with its much superior performance over some state-of-the-art FedC algorithms, and the consistency with all the presented analytical results.
\\\\
\end{abstract}

\ifconfver \else \IEEEpeerreviewmaketitle} \fi

\vspace{-0.8cm}

\section{Introduction}
\begin{figure}[t]
\begin{center}
\resizebox{0.75\linewidth}{!}{\hspace{-0cm}\includegraphics{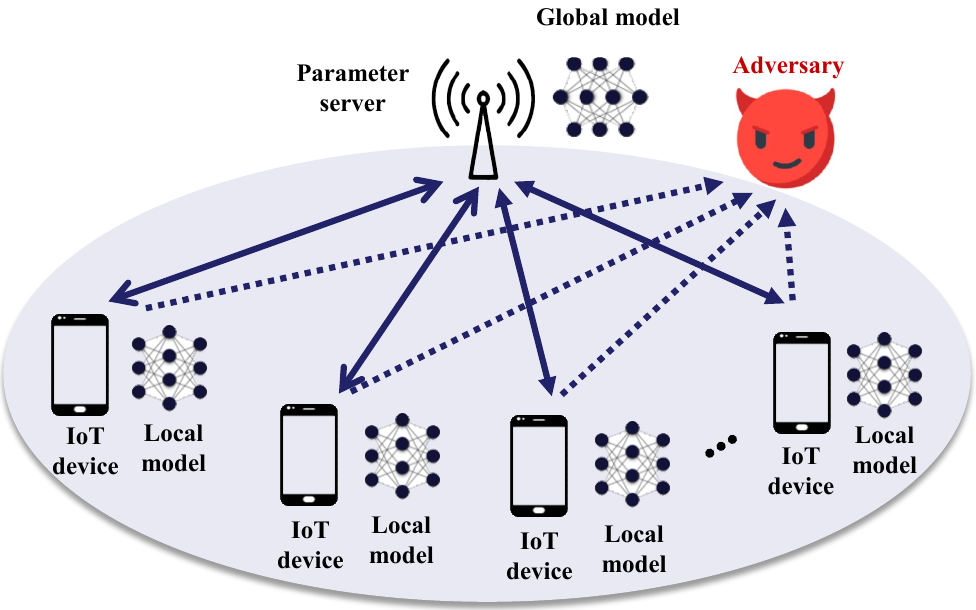}}
\end{center}
\vspace{0.02cm}
\caption{The framework of FL system in the presence of adversaries.}
\label{fig:fedavg_1}
\end{figure}

\IEEEPARstart{F}{ederated} learning (FL), as a novel distributed paradigm,  enables massively distributed clients to jointly find a desired model through machine learning (ML) under the orchestration of a parameter server (PS) while refraining the clients' sensitive data from being exposed~\cite{wang2021federated,McMahan2016FederatedLO}. FL has received tremendous attention in the past several years as it seriously takes numerous practical challenges into account, including limited communication resources and data heterogeneity and client privacy protection in the learning process~\cite{li2022federated}. Under these challenges, most FL algorithms follow a computation-aggregation protocol by which the local update of model parameters and the PS aggregation are repeated in a round-by-round fashion until convergence. Federated average (FedAvg) algorithm~\cite{CE_DDNN_2017,li2022secure} is a typical one,  which improves  communication efficiency by adopting partial client participation (PCP) and multiple  local  stochastic gradient descent (local SGD) updating steps. Nevertheless, the data heterogeneity (e.g., non-identically and independently distributed (non-i.i.d.) data) has been acknowledged to be the main bottleneck to FL deployment. Numerous efforts have been devoted to the analysis of the adverse effects of non-i.i.d. data on algorithm convergence~\cite{CE_DDNN_2017,Parallel_RSGD_2019}.  In parallel, FL still suffers from privacy leakage as the clients' sensitive information could be inferred by adversaries through the exchanged model parameters between the clients and the PS~\cite{bagdasaryan2020backdoor, geiping2020inverting,wei2021gradient}. As illustrated in Fig. \ref{fig:fedavg_1}, a vanilla FL system  includes  many clients and one PS, where the uploaded parameters from local clients may be overheard by  powerful adversaries.
The differential privacy (DP) technique has recently gained increasing popularity in enhancing  privacy of FL thanks to its algorithmic simplicity, support by rigorous mathematical theory, and negligible system overheads~\cite{li2020secure, dwork2014algorithmic}.
Despite the recent rapid progress of FL,  substantial attention has been given to supervised learning, whereas the problem of federated unsupervised learning, especially data clustering, has not yet been investigated comprehensively in FL community~\cite{wang2021clustering}. Clustering in the FL setting, called federated clustering (FedC), aims to partition data samples distributed over massive clients based on a global similarity measure while keeping them on respective clients.
As clustering is one of most suitable missions for ML  and has a great deal of  applications, the FedC and its implementation is believed to be in  impending need. On the other hand, recent years have witnessed an incessant springing up of FedC applications, which again motivates research efforts in this direction. For example, in e-commerce applications, FedC is widely used to group the online customers of multiple institutions with sensitive features, such as personal details, purchase orders, and bank transaction records, to identify their specific interests for precise service recommendation~\cite{stallmann2022towards,kolluri2021private}.
Note that, FedC is quite different from clustered federated learning approaches~\cite{ghosh2019robust,Sattler20}, which, instead of data clustering, are concerned with clusters of clients such that each cluster comes up with a local model to be uploaded to the PS in order to reduce the communication cost of supervised FL systems~\cite{fraboni2021clustered,Dennis21a,ma2022convergence}.

In this paper, an effective FedC algorithm is proposed, that considers both  non-i.i.d. data and DP-based privacy protection. In FedC scenarios where data heterogeneity is prevalent, the global cluster information may not be available for each client as all the data in hands may belong to just a few clusters, and the correct cluster structure might become apparent only when the local datasets are combined~\cite{ma2022convergence}. Moreover, effectively transferring the centralized clustering algorithms into FedC, such as $k$-means, is almost formidable due to the privacy concern. Directly applying them to FedC by following the computation-aggregation protocol would result in serious  performance degradation~\cite{wang2021clustering,ding2016k}. In addition, different from supervised FL, the process of FedC involves the iterative constrained optimization of both cluster centroids and cluster assignments of all data samples, which again brings more difficulties to algorithm design.

As for privacy protection, such coupling optimization necessitates a more careful and fine-grained design and analysis of the DP-based FedC algorithms. In particular, it is widely known that DP protects privacy at the cost of learning performance loss~\cite{dwork2006our}, and balancing the tradeoff between protection level and convergence performance,   so-called the privacy-utility tradeoff, is essential in practical FL applications. To improve the privacy-utility tradeoff,  privacy amplification~\cite{erlingsson2019amplification,balle2018privacy} has been pervasively adopted in many DP-based FL (DP-FL) applications~\cite{li2022secure,shen2022performance}. The privacy amplification can reduce the variance of noise added to locally uploaded models without sacrificing the privacy protection level, thereby mitigating the adverse effects of DP~~\cite{li2022network}.  In addition to the challenges posed by non-i.i.d. data and privacy protection, the practical application of FedC algorithms in  FL systems requires careful consideration of communication cost and straggler effect~\cite{li2019convergence}. These factors and concerns not only affect the algorithm design, but also make the associated theoretical algorithm performance analysis much more involved.
However, the involvement of cluster centroids and data's cluster-membership assignment  in FedC further complicates the design of DP, and it is still not clear how to achieve a good privacy-utility  tradeoff in FedC.

\subsection{Related works}\label{subsec: related work}
Currently,  many successful methods have been reported about traditional distributed clustering, however, they are simply parallel implementations of the centralized clustering algorithms \cite{Kmeans_2012,PKmeans_2014,DBSCAN_1996} or implementations through clustering representative data samples collected from distributed clients~\cite{DKCoreset_2013,Dis_kmeans_2016}. Apparently, the critical challenges of FL, such as massive clients, limited communication resources and data heterogeneity, were rarely  considered, and the demand for privacy protection was also overlooked.

The recent works in \cite{stallmann2022towards, pedrycz2021federated,hernandez2021federated,li2021federated} have considered the FL scenarios and presented FedC algorithms, while most of them were developed by combining the simple centralized $k$-means algorithm (and its variants) with federated average (FedAvg)~\cite{CE_DDNN_2017}. Specifically, in each communication round, the clients employ $k$-means algorithms to obtain the local cluster centroids, which are then uploaded to the PS to produce the global clusters.
The works~\cite{stallmann2022towards, pedrycz2021federated} adopted the fuzzy $k$-means
to perform local clustering, while the global centroids are obtained from the received local centroids by $k$-means clustering.  The work~\cite{hernandez2021federated} proposed a federated spectral clustering approach to train a generative model for each cluster, such that each data sample can be classified to only one cluster using the generated models.
Nevertheless,   the above-mentioned  FedC algorithms did not consider the data heterogeneity issue, thus hardly yielding satisfactory clustering performance as the clustering algorithm only works well in clustering datasets that are evenly spread around the centroids but fails in clustering datasets of complex and heterogeneous cluster structure~\cite{ma2022convergence,Dis_kmeans_2016,xu2021asynchronous}.

To the best of our knowledge, only few works have specifically addressed federated clustering in the context of non-i.i.d. data~\cite{Dennis21a,wang2022federated,chung2022federated,ghosh2019robust}. However, these works also have their limitations.
The approach  reported in~\cite{ghosh2019robust} directly apply the conventional $k$-means to the FL framework, resulting in suboptimal clustering performance, that will be discussed in Section~\ref{sub: clustering}.
The work~\cite{Dennis21a} considered one-shot FedC, where each client obtains a local model using $k$-means and then upload the trained model only once for the aggregation by the PS. However, the one-shot FedC may not be very effective when the FL problem under consideration is NP-hard or non-convex due to the low-quality of local solutions.
The work~\cite{chung2022federated} proposed a federated clustering scheme by assuming that heterogeneous data to be clustered come  from a probabilistic model, that is, assign each data point to a cluster model  with the highest likelihood. More importantly, these works~\cite{Dennis21a,ghosh2019robust,chung2022federated} lack a complete theoretical analysis of the impact of non-i.i.d. data on convergence performance.   The work in~\cite{wang2022federated} formulates the clustering problem as a constrained non-convex problem and theoretically analyze the impact of non-i.i.d. data. However, none of above-mentioned works ever consider the crucial issue of privacy protection, which we believe, is one of the most fundamental concerns in FL system.
The work~\cite{li2022secure} is the first that adopted a secret sharing approach to protect privacy in the federated $k$-means algorithm. However, such a strategy requires complicated encryption protocols and substantial extra communication and computation cost~\cite{bonawitz2017practical}, thus not applicable to large-scale FL models. As far as we are aware,  none of the existing works simultaneously consider data heterogeneity and privacy protection, hence motivating us to develop advanced privacy-preserving federated clustering algorithms over non-i.i.d. data.

\subsection{Contributions}
Motivated by the aforementioned issues of existing FedC methods, we propose a differentially private FedC algorithm, called DP-FedC, with the data heterogeneity and privacy protection taken into account.
The main contributions of this work are summarized as follows:
\vspace{-0.0cm}
\begin{enumerate}

\item A  novel clustering  problem is formulated to overcome the shortcomings of the conventional centralized $k$-means, then applied it to  FedC scenarios.
To handle the proposed FedC problem, a DP-FedC algorithm under the computation-aggregation protocol is developed, that alternatively update local cluster centroids and indicator matrices (indicating each sample and the cluster it belongs) through allowing multiple local SGD steps and partial clients participation. Furthermore, the  privacy amplification strategy is employed to reduce the DP noise variance for better tradeoff between learning performance (i.e., clustering accuracy)  and privacy protection.

\item Two theoretical analyses for the proposed DP-FedC algorithm are presented. One is a privacy analysis, showing that a tighter upper bound of the total privacy loss, i.e.,  $(\mathcal{O}(q\epsilon \sqrt{p R}), \delta)$-DP over $R$ consecutive communication rounds, where $0<p, q \leq 1$ are defined in Remark~\ref{remark:remark_dp}.  The other is a convergence analysis, showing the convergence rate $\mathcal{O}(1/\sqrt{R})$ under non-convex and   non-i.i.d. data setting.

\item Extensive experimental results are provided to demonstrate the effectiveness of the proposed DP-FedC algorithm on real world datasets, including TCGA cancer gene data~\cite{TCGA_CGCD}, and the MNIST hand-writing digits data~\cite{website_MNIST}, and its much superior performance over state-of-the-art distributed clustering and FedC algorithms.
\end{enumerate}

\vspace{-0.0cm}
{\bf Synopsis:}   Section \ref{sec: Preliminaries}     introduces    some  preliminaries of DP.  Section \ref{sec: problem formulation}  presents the  problem formulation.
Section \ref{sec: fedam}   presents  the proposed  DP-FedC algorithm. Section \ref{sec:privacy_analysis}  focuses  on privacy analysis and convergence analysis of the proposed algorithm. Experiment results are presented in Section \ref{sec: simulation}, and finally the conclusion is drawn in Section \ref{sec: conclusion}.

{\bf Notation:}
$\E[\cdot]$ represents the expectation of random variables or events;  $\operatorname{Pr} [\cdot]$ represents the   probability function;
$\mathbb{R}^{m \times n}$ denotes the set of $m$ by $n$ real-valued matrices;
The $(i,j)$-th entry of matrix ${\bf A}\in{\mathbb R}^{m\times n}$ is denoted by $\Ab (i,j)$;   $\Ab(i,:)$ and $\Ab(:, j)$ denote the $i$-th row and the $j$-th column of $\Ab$, respectively;  $\Ab \geq 0$ means $\Ab (i,j)\geq 0, \forall i,j$;
$[\Ab]^+$ denotes the matrix by replacing all the negative elements in $\Ab$ with zero.
$\lambda_{\max}(\Ab)$ stands for the maximum eigenvalue of $\Ab$;
$\|\cdot\|_F$,    $\|\cdot\|$ and $\|\cdot\|_0$ are the matrix Frobenius norm, Euclidean norm (i.e., $\ell_2$-norm) and zero norm of vectors,  respectively;
$\langle \xb,   \yb \rangle = \xb^{\top} \yb $ represents the inner product operator, where the superscript $`\top$'  denotes the  vector transpose;
For any integer $N$, $[N]$  denotes the integer set $\{1,\ldots, N\}$;
$\oneb$ denotes the all-one vector; 
$\{\Cc_i \}_{i=1}^{k}$ denotes   the set $\{{\cal C}_1,{\cal C}_2\dots,{\cal C}_k\}$;
$\lfloor \cdot \rfloor$ denotes the floor function.

\section{Preliminaries}\label{sec: Preliminaries}
\subsection{Differential privacy}\label{subsec:Privated FedONMF}
In this work, we assume that any third party is untrustworthy, including the honest-but-curious server.
The core privacy protection mechanism of the proposed DP-FedC is the well-known DP based random mechanism defined as follows:
\vspace{0.1cm}
\begin{Def}\label{Def:DP defintion}  $(\epsilon, \delta)$-DP {\rm \cite{dwork2014algorithmic}}.
Consider two neighboring datasets  $\mathcal{D}$ and $\mathcal{D}^{\prime}$, which differ in only one data sample. A randomized mechanism $\mathcal{M}$ is $(\epsilon, \delta)$-DP if for any two $\mathcal{D}$, $\mathcal{D}^{\prime}$ and measurable subset $\Oc \subseteq Range(\mathcal{M})$,
\begin{align} \label{eqn:DP_def}
\operatorname{Pr}[\mathcal{M}(\mathcal{D}) \in \Oc] \leq \exp(\epsilon) \cdot \operatorname{Pr}\left[\mathcal{M}( \mathcal{D}^{\prime}) \in \Oc \right]+\delta,
\end{align}
holds true,  $\epsilon > 0$ represents the privacy protection level,  and $0< \delta < 1$ is the probability threshold to break $(\epsilon, 0)$-DP.
\end{Def}
A smaller $\epsilon$ means  that it is more difficult to distinguish between the two neighboring datasets $\mathcal{D}$ and $ \mathcal{D}^{\prime}$, thus resulting in stronger privacy protection.
The required ``noise variance" $\sigma^2$  for achieving $(\epsilon, \delta)$-DP  is given by the following lemma.
\begin{Lemma}  \label{Lemma: global sensitivity}
\cite[Theorem 3.22]{dwork2014algorithmic}   Suppose a query function $g$  accesses the dataset $\Dc$ via randomized mechanism  $\mathcal{M}$.
Let $\xi$ be a zero-mean Gaussian noise with variance $\sigma^2$. Then $g+ \xi$ is $(\epsilon,\delta)$-DP if
\begin{align}
\sigma^2 = \frac{2 s^{2} \ln(1.25/\delta)}{\epsilon^{2}},
\end{align}where $ s$  is  the  $\ell_2$-norm sensitivity  of the function $g$ defined by
\begin{equation}\label{eqn:global sensitivity_f}
\begin{aligned}
s \triangleq \max _{\mathcal{D}, \mathcal{D}^{\prime}}\big\|g(\mathcal{D})- g\big(\mathcal{D}^{\prime}\big)\big\|.
\end{aligned}
\end{equation}
\end{Lemma}

In practical  FL systems, it is crucial to monitor the total privacy loss over multiple communication rounds of model parameters exchange with the PS, which can be computed from the individual privacy loss stated in the following definition.
\begin{Def}\label{Def:Total privacy loss} (Privacy loss~\cite{dwork2014algorithmic}). Suppose that a randomized mechanism $\mathcal{M}$ satisfies $(\epsilon, \delta)$-DP. Let ${\cal D}$ and ${\cal D}^\prime$ be two neighboring datasets and   $\ob$ be a possible random vector of ${\mathcal{M}} (\Dc)$ and ${\mathcal{M}} ({\Dc}^{\prime})$. Then, the privacy loss    is  defined by
\begin{align} \label{eqn:def2}
\rm{PL}(\ob ) \triangleq \ln \Big(\frac{ \mathbb{P}\big[ {\mathcal{M}} (\Dc)= \ob  \big]}{\mathbb{P}\big[\mathcal{M}( {\Dc}^{\prime})= \ob \big]}\Big).	
\end{align}
\end{Def}
Note that, the computation of total privacy loss is quite involved, though its upper bound can be estimated using the moment accountant method~\cite{abadi2016deep},  which so far yields the tightest bound on the total privacy loss.

According to the privacy amplification theorem~\cite{balle2018privacy}, it has been known that, running on a randomly generated subset of a
dataset, the DP mechanism can yield stronger privacy protection than running on the entire dataset. This fact implies that the noise variance required
for achieving a predefined DP level can be reduced when partial data are randomly selected at each iteration.
The privacy analysis to be addressed in Section \ref{subsec:privacy_analysis} relies on the following privacy amplification theorem.
\begin{Theorem} \label{thm: privacy amplicfication via sampling}
(Privacy Amplification Theorem~~\cite{balle2018privacy}) Suppose that a mechanism $\mathcal{M}$ is $(\epsilon, \delta)$-DP over a given dataset $\mathcal{D}$ with size $n$. Consider the subsampling mechanism that outputs a random sample uniformly over all subsets $\mathcal{D}_{s} \subseteq \mathcal{D}$ with size $b$. Then, when $\epsilon \leq 1$,  executing $\mathcal{M}$  mechanism   on the subset $\mathcal{D}_{s}$ guarantees $(\epsilon^{\prime},  \delta^{\prime})$-DP, where $\epsilon^{\prime}$ and $\delta^{\prime}$ are given by
\begin{align}
\epsilon^{\prime} = \min (2q \epsilon,\epsilon), \delta^{\prime}=q \delta,
\end{align}where $q=b/n$ is the data sampling ratio.
\end{Theorem} 	
\textit{Proof:}  See Appendix \ref{appendix:proof Theorem1}. \vspace{0.1cm}

According to Theorem \ref{thm: privacy amplicfication via sampling}, the privacy would be amplified when $q \leq 1/2$.  Note that, the privacy amplification for local DP is pervasively adopted in existing FL literatures~\cite{li2020secure,shen2022performance} since only a small portion of data being used in local SGD.
\vspace{-0.3cm}
\subsection{Centralized $k$-means clustering}\label{sub: clustering}
Let $\Xb  $ be a  data matrix that contains $n$ data samples and each sample has $m$ features, i.e., $\Xb =$ $\left[\xb_{1}, \ldots, \xb_{n}\right] \in \Rbb^{m \times n}$. The clustering task is to assign the $n$ data samples of $\Xb$ to a predefined number of $k$ clusters such that the samples within a cluster are closer to each other than to those belonging to any other cluster in terms of a certain distance metric. Among hundreds of clustering algorithms, the most classic and popular one is the $k$-means which aims to obtain $k$ non-overlapping clusters $\{\Cc_i \}_{i=1}^{k}$, i.e.,  $\Cc_i \cap \Cc_{i^{\prime}}=\emptyset, \forall i \ne i^\prime \in [k], \bigcup_{i=1}^{k} \Cc_i = \{\xb_{j}\}_{j=1}^{n}$,  by minimizing the average Euclidean distance between each cluster centroid and all the data samples within the cluster.

From the optimization perspective, the $k$-means algorithm can be viewed as an ad hoc algorithm, which handles the following matrix factorization by alternative minimization (AM)~\cite{yang2016learning}:
\begin{subequations}\label{eqn: kmeans_prob}
\begin{align}
\min_{\substack{\Wb \in{\mathbb R}^{m \times k}, \Hb  }}~&
\|\mathbf{X}-\mathbf{W H}\|_{F}^{2}\\
{\rm s.t.}~& \Hb \in \{0,1\}^{k \times n}, \|\Hb(:,j)\|_0 =1, \forall j, \label{eqn: prob_H_con}
\end{align}
\end{subequations}
where    ${\bf W}\in {\mathbb R}^{m\times k}$ is a matrix consisting of the $k$ centroids, and ${\bf H} \in {\mathbb R}^{k\times n}$ is an indicator matrix with only one non-zero element (i.e., unity) in each column.
Applying AM to problem \eqref{eqn: kmeans_prob} gives rise to the following update rules of $\Wb$ and $\Hb$ at   iteration $t+1$:
\begin{align}
\Hb^{t+1} =& \arg\min_{\bf{H}}~\|\Xb - \Wb^{t} \Hb \|_F^2, \notag \\
&~~{\rm s.t.}~ \Hb \in \{0,1\}^{k \times n}, \|\Hb(:,j)\|_0 =1, \forall j. \label{eqn: kmeans_W_update}\\
\Wb^{t+1} = &\arg\min_{\Wb \in{\mathbb R}^{m \times k} }~\|\Xb - \Wb\Hb^{t+1}\|_F^2. \label{eqn: kmeans_H_update}
\end{align}
Closed-form solutions to \eqref{eqn: kmeans_W_update} and \eqref{eqn: kmeans_H_update} are respectively given by
\begin{small}
\begin{align}
&\Hb^{t+1}(l, j) = \begin{cases}
1 & {\rm if~}l= \arg\min\limits_{u} \|\Xb(:, j) - \Wb^t(:, u)\|^2, \\
0 & \text{otherwise},
\end{cases}
\end{align}
\end{small}and
\begin{small}
\begin{align}
\Wb^{t+1}(:,l) = \frac{1}{|\Jc_l^t|} \sum_{u \in \Jc_l^t} \Xb(:,u),
\end{align}
\end{small}where $\Jc_l^t = \{ j| \Hb(l; j) = 1 \}$. Note that at iteration $t+1$, the $l$-th row of ${\bf H}$ is updated according to the minimum distance from each data sample to the $l$-th centroid according to ${\bf W}^t$, and then the $l$-th centroid (i.e., the $l$-column of ${\bf W}$) is updated as the average of the data belonging to cluster $l$ according to the $l$-th row of the updated indicator matrix.

\section{Problem Formulation}\label{sec: problem formulation}
The centralized $k$-means may totally fail for the dataset with complex distribution and data heterogeneity, so it is not suitable  for distributed environments, especially the FL setting.
The reasons are twofold. First, the non-convex $k$-means problem~\eqref{eqn: kmeans_prob} is NP-hard due to involving binary variables, and hence almost any algorithm (including $k$-means) is unable to work well.
No wonder, it's performance is quite sensitive to the initial conditions, complex data distribution, and the obtained solution easily trapped in bad local minima and so forth~\cite{K++_2007}.
Moreover, the less data samples the worse its performance, thus further downgrading its performance in FL scenarios, especially when the data are sensitive and under privacy concern.
Most existing FedC algorithms are based on $k$-means and operate in computation-aggregation fashion, but their performance may get seriously downgraded under FL scenarios, including massively distributed clients and severe client heterogeneity~\cite{FedProx_2018}.

Inspired by the idea in \cite{Shuai_SNCP_2019}, we replace the binary constraint \eqref{eqn: prob_H_con} with a norm-based equality constraint and reformulate problem~\eqref{eqn: kmeans_prob} as
\begin{subequations}\label{eqn: cluster_prob1}
\begin{align}
\min _{\mathbf{W}, \mathbf{H}} &~  \|\Xb-\Wb \Hb\|_{F}^{2}  +   \frac{\mu_{h}}{2} \|\Hb\|_{F}^{2}  + \frac{\mu_{w}}{2}\|\Wb\|_{F}^{2} \label{eqn: cluster_prob1_ca}\\
\text { s.t. } &\Hb \geq 0, \left\|\Hb(:,j)\right\|_{1}^{2}=\left\|\Hb(:,j)\right\|_{2}^{2}, \forall j \in [n], \label{eqn: cluster_prob1_c}
\end{align}where $\mu_{h} > 0$ and $\mu_{w} > 0$ are two positive parameters.
\end{subequations}
Problem \eqref{eqn: cluster_prob1} is a non-convex and non-smooth problem and it can be regarded as a relaxation of the $k$-means problem \eqref{eqn: kmeans_prob} because ${\bf H}$ has been relaxed as a real $k\times n$ matrix, with at most one non-zero entry (not equal to one) in each column, though the equality constraint \eqref{eqn: cluster_prob1_c} is still non-convex.  Moreover, the two   regularization terms (i.e., the 2nd and the 3rd terms in \eqref{eqn: cluster_prob1_ca}) are used to control the resulting scaling/counter-scaling ambiguity \cite{yang2016learning}.

Instead of directly solving problem \eqref{eqn: cluster_prob1}, we consider the following problem by dropping the equality constraint in \eqref{eqn: cluster_prob1_c} and adding an associated penalty term in the objective function:
\begin{subequations}\label{eqn: cluster_prob2}
\begin{align}
\min _{\Wb, \Hb}  ~&  \|\Xb-\Wb \Hb\|_{F}^{2} +   \frac{\mu_{h}}{2} \|\Hb\|_{F}^{2}  + \frac{\mu_{w}}{2}\|\mathbf{W}\|_{F}^{2} + \frac{\rho}{2} \sum_{j=1}^{n}\left(\left(\mathbf{1}^{\top} \Hb(:,j) \right)^{2}-\left\|\Hb(:,j) \right\|_{2}^{2}\right) \\
\text { s.t. } &  \mathbf{H} \geq 0, \label{eqn: cluster_prob2_c2}
\end{align}
\end{subequations}where $\rho > 0$ is a penalty parameter. The larger the value of $\rho$, the smaller the approximation error of the equality constraint in \eqref{eqn: cluster_prob1_c} and the more sparse the matrix ${\bf H}$.
It is remarkable that   problem \eqref{eqn: cluster_prob2} is much efficient to handle than problem \eqref{eqn: cluster_prob1} for two reasons. One is that \eqref{eqn: cluster_prob2_c2} is a simple convex constraint; the other is that the assignment of each data sample to an unique cluster is not reliable for problem \eqref{eqn: cluster_prob1}~\cite{SNMF_Kim,Zhang_ONMF_SM2016}.
Therefore, in contrast to the hard clustering performed by $k$-means, solving \eqref{eqn: cluster_prob2} corresponds to seeking a soft clustering solution~\cite{yu2005soft} instead.

\begin{figure}[t]
\begin{center}
\resizebox{0.75\linewidth}{!}{\hspace{-0cm}\includegraphics{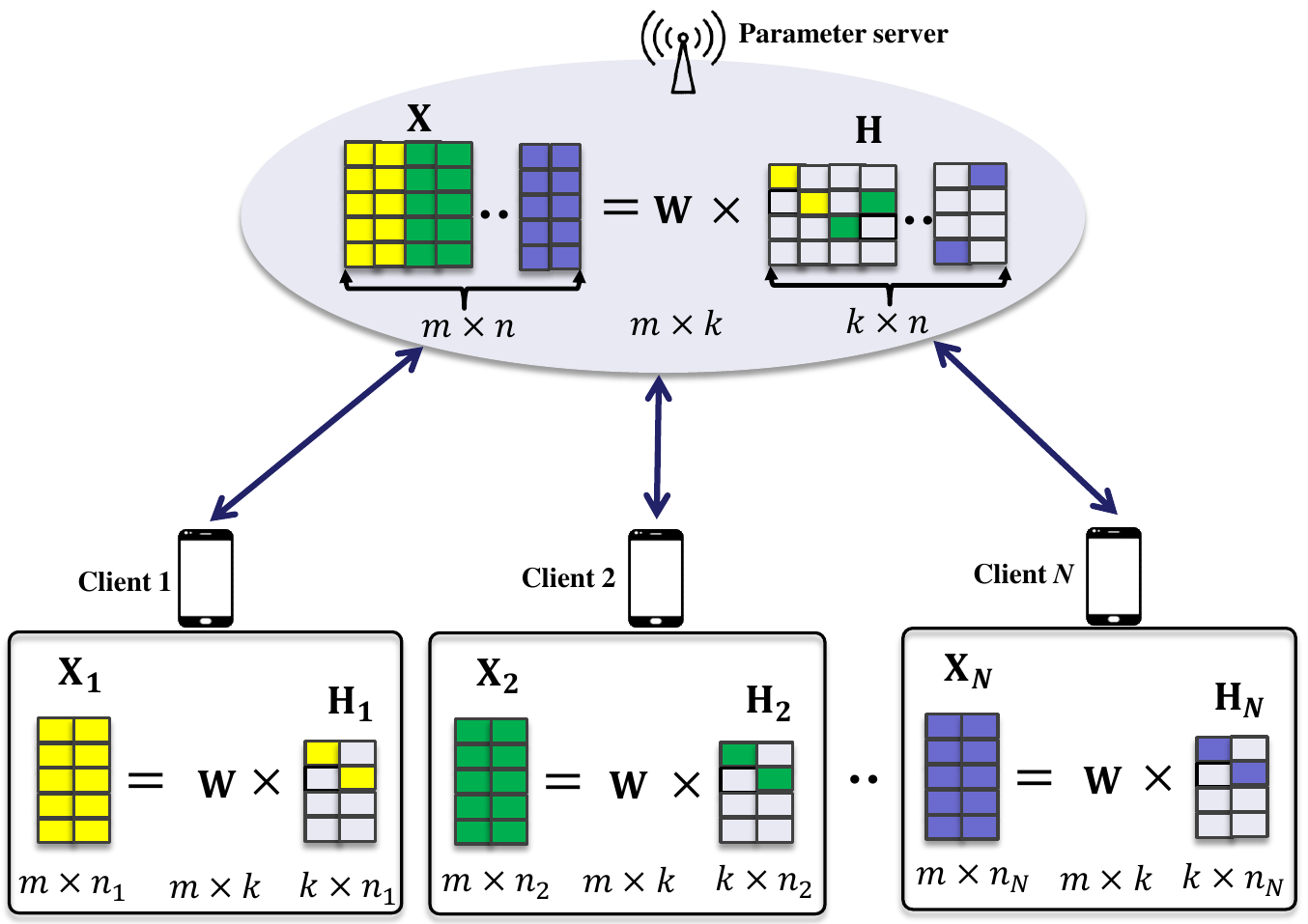}}
\end{center}
\vspace{0.05cm}
\caption{The proposed framework for federated clustering.}
\label{fig:fed_clustering}
\end{figure}
\vspace{0.2cm}
\subsection{Federated clustering model}
To solve problem \eqref{eqn: cluster_prob2}  under the FL network,  we first assume that the data matrix is partitioned and distributed over $N$ clients. i.e.,  $\Xb=[\Xb_1,\Xb_2,\ldots,\Xb_N]$. Specifically, each client $i$ owns non-overlapping data $\Xb_i\in \mathbb{R}^{m \times n_i}$, where $n_i$ is the number of data samples in client $i$ and $\sum_{i = 1}^{N} n_i = n$. Under the FL scenario, $N$ could be large, and the data $\Xb_1,\Xb_2,\ldots,\Xb_N$ could be unbalanced and non-i.i.d. \cite{FL_Beyond_2015,FL_Ondevice_2016}.
We proceed  by partitioning   $\Hb$  in the same fashion as $\Xb$, resulting in the form $\Hb=[\Hb_1,\Hb_2,\ldots,\Hb_N]$. Each column of  $\Hb$ corresponds to a certain data sample in $\Xb$, while $\Wb$ is treated as shared parameters that will be uploaded to the PS for information exchange.  The resulting framework for federated clustering is illustrated  in Fig.~\ref{fig:fed_clustering}, where a central PS coordinates the $N$ clients to accomplish the clustering task.
Then, one can reformulate problem \eqref{eqn: cluster_prob2} into the FL framework as follows.
\begin{subequations}\label{eqn: Fed_cluster_prob}
\begin{align}
\min_{\substack{\Wb,~ \Hb_i, \\
i = 1, \ldots, N}} ~&F(\Wb, \Hb)\triangleq \frac{1}{N}\sum_{i = 1}^{N}  F_i(\Wb, \Hb_i) \\
{\rm s.t.}~&\Hb_i \geq 0, \forall i \in [N],
\end{align}
\end{subequations}where
\begin{small}
\begin{align}\label{eqn: obj of client p}
\hspace{-0.0cm} F_i(\Wb, \Hb_i) \triangleq & \|\Xb_i-\Wb \Hb_i\|_{F}^{2}  + \frac{\rho}{2}( \tr(\Hb_{i}\Ub \Hb_{i}^\top)-\|\Hb_i\|_F^2) + \frac{\mu_{h}}{2} \|\mathbf{H}_i\|_{F}^{2}  + \frac{\mu_{w}}{2}\|\mathbf{W}\|_{F}^{2}
\end{align}
\end{small}is the local objective function of each client $i$, and $\Ub \triangleq \oneb\oneb^\top$.

In contrast to the vanilla FL problem which contains only one shared optimization variable, problem \eqref{eqn: Fed_cluster_prob} involves two  variables: one is $\Wb$ which is the cluster centroid matrix $\Wb$ shared among clients, and the other one is $\Hb_{i}$ which is local cluster indicator matrix for $\Xb_i$ owned by client $i$. This apparently brings challenges in the algorithm development, especially in the presence of non-i.i.d. data. In parallel, as $\Wb$ is shared, there certainly exist possibilities of leaking clients' privacy in the FL process. Recent work~\cite{SFMF_2019} showed that the honest-but-curious server could infer clients'  private data from the uploaded information in the federated matrix factorization framework. Consequently,  it is inevitable to develop  an effective and privacy-preserving FL algorithm for problem \eqref{eqn: Fed_cluster_prob}.

\section{Proposed algorithm for problem \eqref{eqn: Fed_cluster_prob}} \label{sec: fedam}
In this section, we develop a novel FedC algorithm to solve~\eqref{eqn: Fed_cluster_prob},   which judiciously updates $\Wb$ and $\Hb_i, i \in [N]$, and adopts an   amplified DP for rigorous privacy protection.

\subsection{Update of $\Wb$ and $\Hb_i$ in FL}\label{sec: fedam development}
The key of algorithmic development to problem \eqref{eqn: Fed_cluster_prob} is to specify how to perform the local update of $\Hb_{i}$ and global update of $\Wb$. Inspired by \cite{li2019convergence}, we follow the same spirit of local SGD and PCP, where a subset of clients are selected to locally update $\Hb_{i}$ and the associated local copies $\Wb_{i}$'s of $\Wb$, and then upload these iterates to the PS for global aggregation in each round. In particular, for round $t=1,2,\ldots$,
\begin{enumerate}[(a)]
\item \textbf{Client sampling}:  We let the PS uniformly sample a small and fixed-size set $\Sc^{t}$ of $K$ clients, i.e., $\Sc^{t} \subseteq [N],  |\Sc^t|=K$, and then broadcast the global $\Wb^{t-1}$ to all clients.

\item \textbf{Local  update}:  All clients are asked to obtain an approximate solution $(\Wb_i^t, \Hb_i^t)$ to the following local subproblem of \eqref{eqn: Fed_cluster_prob}.
\begin{align} \label{eqn: local_irob}
(\Wb_i^t, \Hb_i^t ) =  \arg\min_{\substack{\Wb, \Hb_i \geq 0}} ~&F_i(\Wb, \Hb_i).
\end{align}
After that, each client $i \in \Sc^t$ uploads $\Wb_i^{t}$ to the PS.
\item \textbf{Global aggregation}: After receiving $\Wb_i^{t}$ from all clients $i \in \Sc^t$, the PS aggregates them to produce the new global $\Wb^t$, i.e.,
\begin{align} \label{eqn: FedAM proj of W PCC}
\Wb^t = \frac{1}{K} \sum_{i \in \Sc^{t}} \Wb_i^{t}.
\end{align}
\end{enumerate}

In order to specify the local iterates $(\Wb_i^t, \Hb_i^t )$, we propose to handle  \eqref{eqn: local_irob} by combining AM~\cite{tseng2001convergence} and local SGD.  That is,
$\Hb_i^t$  is produced by applying multiple gradient descent (GD) steps to \eqref{eqn: local_irob} with $\Wb_i$     fixed, and then $\Wb_i^t$ is updated similarly by fixing  $\Hb_i$. To be more specific, we first let all clients perform $Q_1 \geq 1$ consecutive steps of projected GD with respect to $\Hb_i$, i.e.,  for $r=1, \ldots, Q_1$,

\begin{align}\label{eqn: FedAM update of H1 0}
\hspace{-0.1cm}	& \Hb_i^{t, r}\! =\!  \Big[ \Hb_i^{t,r-1}\! -\! \frac{1}{\gamma_i^{t}}{\nabla_{H_i}F_i(\Wb^{t-1}, \Hb_i^{t,r-1})} \Big]^{+},
\end{align}where  $\gamma_i^{t}>0$ is the learning rate. Then, they are asked to perform $Q_2^t \geq 1$ consecutive steps of SGD (no projection) with respect to $\Wb$, i.e., for $r =  Q_1 + 1, \ldots, Q^t,$
\begin{align}\label{eqn: FedAM update of W2 0}
\hspace{-0.1cm}&\Wb_i^{t, r} = \Wb_i^{t, r-1} - \frac{1}{\eta^t}{\nabla_{W}F_i(\Wb_i^{t, r-1}, \Hb_i^{t, Q_1};\mathcal{B}_i^{t,r})},
\end{align}where $Q^t=Q_1+Q_2^t$ and $\eta^t>0$ is a step size, and $\nabla_{W}F_i(\Wb_i^{t, r-1}, \Hb_i^{t}; \mathcal{B}_i^{t,r})$  is the  stochastic gradient computed using mini-batch dataset $\mathcal{B}_i^{t,r}$  with   size $b$ ($|\mathcal{B}_i^{t,r}|=b$).  Lastly, $(\Wb_i^t, \Hb_i^t )$ is obtained by setting $\Hb_i^{t}=\Hb_i^{t, Q_1}$ and $\Wb_i^t = \Wb_i^{t,Q^t}$.

\subsection{Privacy concern}
Data   privacy is one of primary concerns  in FL systems.  To enhance data privacy, we apply the DP technique to the proposed algorithm. Specifically, in each round $t$, we add an artificially
Gaussian noise matrix ${\xib}_i^t \in {\mathbb R}^{m\times k}$ to ${\bf W}_i^t$, where all the $mk$ entries of ${\xib}_i^t$ are i.i.d. Gaussian random variables with zero mean and variance $\sigma_{i,t}^2$, thus yielding
\begin{align}\label{eqn:model_perturbed}
\wt{\Wb}_i^{t}  =\Wb_i^{t} + \xib_{i}^{t},
\end{align}
and then upload $\wt{\Wb}_i^{t}$ to the PS. Then, \eqref{eqn: FedAM proj of W PCC} becomes
\begin{align}\label{eqn: FedAM proj of W PCC_pertubed}
\Wb^{t+1} =\frac{1}{K} \sum_{i \in \Sc^{t}} \wt{\Wb}_i^{t}.
\end{align}The details of the proposed algorithm are summarized in Algorithm \ref{alg: model_avg}. Note that, the diminishing  $Q_2^t=\lfloor \frac{\widehat{Q}}{t}  \rfloor+1$  (line 12) denotes the number of iterations in updating $ \Wb_i^{t,r}$ (lines 13-15)  by \eqref{eqn: FedAM update of W2 0}, where $\widehat{Q}$ is a given constant and the mini-batch dataset $\mathcal{B}_i^{t,r}$ of size $b$ used is further discussed in the following remark:
\begin{Remark} \label{remark:T1}
For lines 13-15 of Algorithm 1, $Q_2^t b$ data samples are obtained from the dataset $\Dc_i$  at each communication round (i.e., the data sampling ratio $q_{i,t}=Q_2^tb/n_i$), and then divided into $Q_2^t$ mini-batch datasets $\Bc_i^{t,r}$   for each inner iteration $r$.
\end{Remark}

It is acknowledged that the   DP noise matrix ${\xib}_i^t$ will   bring about   adverse effects on  algorithm convergence and     learning performance.
However, the performance degradation of Algorithm~\ref{alg: model_avg} will get worse from round to round due to $\Wb$ perturbed by the DP noise
and the coupling of  $\Wb$ and $\Hb$, on one hand. The accumulated DP noise effects will also get worse with $t$ on the other hand. Therefore, Algorithm~\ref{alg: model_avg} is performance-sensitive to the DP noise in a complicated manner, such that   obtaining a satisfactory privacy-utility tradeoff through theoretical analysis becomes more intractable.

Nevertheless, the privacy amplification presented in Theorem~\ref{thm: privacy amplicfication via sampling}, can be utilized to pursue the performance analysis of Algorithm~\ref{alg: model_avg}, in order to find the clue about the variance reduction of the DP noise for guaranteeing $(\epsilon, \delta)$-DP privacy protection level at each round. The details are presented in the next section.

\begin{algorithm}[!t]
\caption{DP-FedC algorithm} 
\begin{algorithmic}[1]\label{alg: model_avg}
\STATE {\bfseries Input:} initial values of $\Wb_1^{0}=\cdots=\Wb_N^{0}=\Wb^{0}$,  initial values of $\{\Hb_i^{0}\}_{i=1}^N$,
$\Sc^0 = \{1, \ldots, N\}$, $R$ and $\widehat Q$.
\FOR{round $t=1$ {\bfseries to} $R$}
\STATE {\bfseries \underline{Server side:}}
\STATE Compute $\Wb^t$ by \eqref{eqn: FedAM proj of W PCC_pertubed}.
\STATE   Uniformly sample a set of clients $\Sc^t \subseteq [N]$, and broadcast $\Wb^{t}$ to all clients.
\STATE {\bfseries \underline{Client side:}}
\FOR{client $i \in [N]$    in parallel}
\STATE Set $\Hb_i^{t, 0} = \Hb_i^{t-1}$ and  $\Wb_i^{t, 0} = \Wb^t$. 
\FOR{  $r = 1$ {\bfseries to} $Q_1$}
\STATE Update $\Hb_i^{t, r}$ by \eqref{eqn: FedAM update of H1 0}, and set $\Wb_i^{t, r}  = \Wb_i^{t, r-1}$.
\ENDFOR
 \STATE Compute $Q_2^t=\lfloor\frac{\widehat{Q}}{t} \rfloor+1$.
\FOR{$r =  Q_1 + 1$ {\bfseries to} $Q^t=Q_1+Q_2^t$}
\STATE Update $\Wb_i^{t, r}$ by \eqref{eqn: FedAM update of W2 0}, and set $\Hb_i^{t, r}=\Hb_i^{t, r-1}$.
\ENDFOR
\ENDFOR
\STATE Set $\Wb_i^{t}=\Wb_i^{t, Q^t}$ and $\Hb_i^{t}=\Hb_i^{t, Q^t}$.
\FOR{client $i \in \Sc^t$    in parallel}
\STATE Compute $ \wt{\Wb}_i^{t}$   by \eqref{eqn:model_perturbed}.
\STATE Upload $ \wt{\Wb}_i^{t}$ to the PS for next round of aggregation.	
\ENDFOR
\ENDFOR
\end{algorithmic}\vspace{-1mm}
\end{algorithm}

\section{Theoretical Analysis}\label{sec:privacy_analysis}
\subsection{Assumptions}
We need the following assumptions to analyze the privacy guarantee and convergence performance of the proposed algorithm.
\begin{assumption}	\label{Ass: bounded_Lipschitz} Each local cost function $F_i$ is continuously differentiable in both $\Wb$ and $\Hb_i$.
That is, $\nabla_{H_i} F_i(\Wb^{t}, \cdot)$ is Lipschitz continuous   with constant $L_{H_i}^t$, and $\nabla_{W} F_i(\cdot, \Hb_i^{t})$ is Lipschitz continuous  with constant $L_{W_i}^t$, i.e., for any $   \Xb, \Yb$,
{\small
\begin{subequations}
\begin{align}
&\| \nabla_{H_i} F_i(\Wb^{t}, \Xb)  -  \nabla_{H_i} F_i(\Wb^{t}, \Yb) \|_{F} \leq  L_{H_i}^t \| \Xb - \Yb \|_F,\\
&\| \nabla_{W} F_i(\Xb, \Hb_i^{t})  -  \nabla_{W} F_i(\Yb, \Hb_i^{t}) \|_F \leq L_{W_i}^t \| \Xb - \Yb \|_F.
\end{align}
\end{subequations}}
\end{assumption}
According to Assumption \ref{Ass: bounded_Lipschitz} and \cite{wang2022federated}, $\nabla_{W} F(\cdot, \Hb^{t})$ is    Lipschitz continuous with a constant $L_W^t = (\sum_{i = 1}^{N}(L_{W_i}^t)^2/N)^{1/2}$, together with upper and lower bounds for  $L_{H_i}^t$ and $  L_{W_i}^t$, i.e.,
\begin{align}
\ol L_W \geq L_{W_i}^t \geq \underline{L}_{W} > 0,~\ol L_H \geq L_{H_i}^t \geq \underline{L}_{H} > 0, \forall i, t.
\end{align}


\vspace{-0.3cm}
\begin{assumption}\label{Ass: bounded_gradient}
All the local cost functions $F_i$ and their gradients are bounded, i.e., for any $i \in [N]$ and $t$,
\begin{align}
\|\nabla_{W} F_i(\Wb, \Hb_i; \mathcal{B}_i)\|_F^2 &\leq G^2, \forall \Wb, \Hb_i \geq 0, \\
F_i(\Wb,\Hb_i) \geq \underline{F} &> {-\infty},\forall \Wb, \Hb_i \geq 0,
\end{align}where $G$ is a constant, and $\mathcal{B}_i \subseteq \Dc_i$ denotes the mini-batch dataset.
\end{assumption}

\begin{assumption}\label{Ass: bounded_Gradient_variance}
For any mini-batch dataset $\mathcal{B}_i^t$ with size $b$ that are randomly sampled from dataset $\Dc_i$,  the following equations hold,
\begin{align}
&\E [  \nabla_{W}  F_i(\Wb_i^t, \Hb_i^t; \mathcal{B}_i^t)] = \nabla_{W} F_i(\Wb_i^t, \Hb_i^t),    \label{eqn:gradient_unbasis} \\
&\E[\|\nabla_{W} F_i(\Wb_i^t, \Hb_i^t) - \nabla_{W}F_i(\Wb_i^t, \Hb_i^t; \mathcal{B}_i^t)\|_F^2]  \leq \frac{\phi^2}{b},  \label{eqn:bounded_gradient_variance}
\end{align}for any $ i\in [N] $ and $t$, where $\phi$ is a constant.
\end{assumption}
\begin{assumption}	\label{Ass: non_iid}
($\zeta$-non-i.i.d. data)  All the local cost functions  $F_i$ (cf. \eqref{eqn: obj of client p}) are $\zeta$-non-i.i.d., namely, the following condition holds,
\begin{small}
\begin{align}
&\|\nabla_{W} F_i(\Wb, \Hb_i) - \nabla_{W} F(\Wb, \Hb)\|_F^2\leq \zeta^2, \forall \Wb, \Hb \geq 0, \label{eqn: grad_var}
\end{align}
\end{small}where $\zeta \geq 0$ is a  constant.
\end{assumption}
By following similar spirits to those in~\cite{zhang2021fedpd,wang2022federated},  an upper bound $\zeta$ is enforced on all the gradients of $F_i$ and $F$ due to the heterogeneity of local data distributions among  clients. This bound   actually reflects the data's non-i.i.d. degree, which has been extensively utilized in the FL community, particularly for handling non-convex FL problems.

\subsection{Privacy analysis}\label{subsec:privacy_analysis}
\subsubsection{Privacy guarantee}
The $\ell_2$-norm sensitivity~\cite{dwork2014algorithmic} of $\Wb_{i}^{t}$ is stated in following Lemma.

\begin{Lemma}\label{lemma:sensitivity}  For any $t \in [R]$ and $i \in [N]$, the $\ell_2$-norm sensitivity of uploaded local model $\Wb_{i}^{t}$  is given by
\begin{align}
s_{i}^{t}= \frac{2 G Q_2^t }{\eta^t}. \label{eqn:sensitivity_s}
\end{align}
\end{Lemma}
\textit{Proof:}  See the Appendix \ref{proof Lemma_sensitivity}.

According to Lemma \ref{Lemma: global sensitivity} and Lemma \ref{lemma:sensitivity}, we further come up with the following theorem, which can serve as a guideline for determining the variance of DP noise  necessary to fulfill  the associated DP-based FL.
\begin{Theorem}\label{Thm:total_noise_variance}
For any client $i \in [N]$, suppose that $\epsilon \leq 1, \delta \leq 1$, and the data sampling ratio $q_{i,t}=Q_2^tb/n_i$ (cf. Remark~\ref{remark:T1}).   Each entry of  $\xib_{i}^{t}$ generated follows   the Gaussian distribution with zero mean and variance $\sigma_{i,t}^2$, where
\begin{align}\label{eqn:noise for with replacement_0}
\sigma_{i,t}^2 =  \frac{32G^2 (Q_2^t)^2 q_{i,t}^2  \ln(1.25q_{i,t}/\delta)}{ (\eta^t)^2 \epsilon^{2}}.
\end{align}
Then each  communication round of the proposed algorithm guarantees $(\epsilon, \delta)$-DP.
\end{Theorem}
\textit{Proof:} In each communication round of the proposed algorithm, each client $i$ performs $Q_2^t$ steps of SGD w.r.t. $\Wb$ by \eqref{eqn: FedAM update of W2 0}, where the mini-batch dataset with size $b$ used is randomly sampled without replacement from local dataset $\mathcal{D}_{i}$. According to   Lemma \ref{Lemma: global sensitivity} and Theorem \ref{thm: privacy amplicfication via sampling}, the  Gaussian noise with variance
\begin{align} \label{eqn:noise_nos}
\sigma_{i,t}^{2} = \frac{2 s_{i,t}^{2} \ln (1.25 / \delta)}{\epsilon^{2} }
\end{align}
can achieve  at least $ (2 q_{i,t} \epsilon , q_{i,t} \delta)$-DP for client $i$, where $q_{i,t}= Q_2^tb/n_i$ is data sampling ratio for client $i$.   Then, by plugging   $s_{i,t}^{2}$ given by \eqref{eqn:sensitivity_s}     into \eqref{eqn:noise_nos}, we obtain
\begin{align} \label{eqn:noise_no_q}
\sigma_{i,t}^{2} = \frac{4G^2 (Q_2^t)^2  \ln (1.25 / \delta)}{(\eta^t)^2 \epsilon^{2} }, \forall i \in [N].
\end{align}

By \eqref{eqn:noise_no_q}, one can achieve an $(\epsilon, \delta)$-DP for $\Wb_i^{t}$,
by replacing $\epsilon$ and $\delta$ in \eqref{eqn:noise_no_q} with $\epsilon/2q_{i,t}$ and $\delta/q_{i,t}$, respectively, thereby leading to \eqref{eqn:noise for with replacement_0}. \hfill $\blacksquare$
%
\subsubsection{Total privacy loss} \label{subsec:total_privacyloss}
As done in~\cite{li2022federated}, we also  use the moments accountant method to estimate the total privacy loss when the algorithm runs   $R$ communication rounds.
\begin{Theorem}\label{Thm:total_budget}
Suppose that the client $i$ is uniformly sampled by the PS with a probability $p_i$  and the data sampling ratio  $q_{i,t}=Q_2^tb/n_i$ (cf. Remark~\ref{remark:T1}), where $Q_2^t=\lfloor\frac{\widehat{Q}}{t} \rfloor+1$.  Then,  with noise variance $\sigma_{i,t}^2$ (stated in Theorem \ref{Thm:total_noise_variance}) used for the generation of the DP noise under $(\epsilon, \delta)$-DP at each communication round, an achievable total privacy loss ${\bar\epsilon}_i$  for  client  $i$ after  $R$ communication rounds is given by
\begin{small}
\begin{align}
\bar{\epsilon}_{i} &= c_{0}  q_{i,t}^2 \epsilon \sqrt{\frac{ p_i R}{1- q_{i,t}} },  \forall i \in [N], \label{eqn:total_privacyloss}
\end{align}
\end{small}where $ c_{0}$ is a constant.
\vspace{-0.1cm}
\end{Theorem}
\textit{Proof:}  The proof basically follows that of Theorem 1 reported in~\cite{li2022federated}. However, we further consider privacy amplification. Thus, the desired result \eqref{eqn:total_privacyloss} can be obtained by replacing the $\epsilon$ with $2 q_{i,t} \epsilon$ in the corresponding $\bar\epsilon_i$ in Theorem 1 of~\cite{li2022federated}. \hfill $\blacksquare$

Theorem \ref{Thm:total_budget} shows that the achievable lower bound  of total privacy loss $\bar{\epsilon}_{i}$ for the proposed DP-FedC is tighter than that of the latest reported in~\cite{li2022federated,abadi2016deep} when $p$ and $q$ are appropriately chosen.
\begin{Remark} \label{remark:remark_dp}
When clients are uniformly  sampled  with a probability $p_i$, by \eqref{eqn:total_privacyloss}  in Theorem \ref{Thm:total_budget}, one can infer that
Algorithm \ref{alg: model_avg} guarantees  $(\mathcal{O}(q\epsilon \sqrt{p R}), \delta)$-DP under $R$ communication rounds, where $p$ and $q$ are given by
\begin{align}
q &= \max_{i,t}\frac{  q_{i,t}^2}{\sqrt{1-  q_{i,t}}}, \forall i \in [N], t \in [R], \\
p &= \max_{i} p_i , \forall i \in [N],
\end{align}
where $q_{i,t}= Q_2^t b/n_i$.
\end{Remark}

\subsection{Convergence analysis} \label{sec:Convergence Analysis}
To  find some convergence conditions, let us define the following sequence
\begin{small}
\begin{align}\label{eqn: q_1}
\ol \Wb^{t, r} =  \left\{\begin{array}{ll}    \frac{1}{K}\sum\limits_{i \in \Sc^t}  \Wb_i^{t, r},         \text { when } r \in  [Q^t-1],   \\
\frac{1}{K}\sum\limits_{i \in \Sc^t} \big( \Wb_i^{t, Q^t} + \xib_i^t \big),      {\text {when }} r = Q^t,   \end{array}\right.
\end{align}
\end{small}
which is actually   the instantaneous weighted average of local models. Motivated by~\cite{wang2022federated}, let

\vspace{-0.4cm}
\begin{small}
\begin{align}
&G_{H}(\ol \Wb^{t,r},\Hb^{t,r}) \triangleq  \sum_{i =1 }^{N}  (\gamma_i^{t})^2 \big\|\Hb_i^{t,r} - \big[\Hb_i^{t,r}   - \frac{1}{\gamma_i^{t}}{\nabla_{H_i} F_i(\ol \Wb^{t,r}, \Hb_i^{t,r})}\big]^{+} \big\|_F^2,~ \forall r\in [Q_1], \label{eqn: prox_H}\\
&G_{W}(\ol \Wb^{t,r}, \Hb^{t, r}) \triangleq       \|{\nabla_{W} F(\ol \Wb^{t,r}, \Hb^{t,r})}\big)\|_F^2, ~\forall r\in [Q^t]\setminus [Q_1]. \label{eqn: prox_W}
\end{align}
\end{small}
\vspace{-0.3cm}

\noindent If $G_{H}(\ol \Wb^{t,r},\Hb^{t,r})= 0$ and $G_{W}(\ol \Wb^{t,r}, \Hb^{t, r})=0$, then $(\ol \Wb^{t,r},\Hb^{t,r})$ is a stationary-point solution of problem \eqref{eqn: Fed_cluster_prob}.
The main theoretical result  for the DP-FedC is the following theorem.

\begin{Theorem} \label{thm: model_avg}
Let $R$ be the total number of communication rounds and $T=RQ_1+ \sum_{t = 1}^{R} Q_2^t$ be the total number of gradient evaluations per client.  Moreover, let $Q_2^t=\lfloor\frac{\widehat{Q}}{t} \rfloor+1$, $\gamma_i^{t} =  \alpha_1 L_H^t/2 $ and $\eta^t = \alpha_2L_{W}^t$,  where $\alpha_1 >1$ and  $\alpha_2 \geq   Q_2^t \big( 3 (1+ \ol L_W^2/\underline{L}_W^2 )\big)^{1/2}$.    Then, under Assumptions \ref{Ass: bounded_Lipschitz}-\ref{Ass: non_iid}, the sequence $\{(\ol \Wb^{t, r}, \Hb^{t, r})\}$ yielded by Algorithm \ref{alg: model_avg} satisfies
\begin{small}
\begin{align}
& \frac{1}{T}\Big[\sum_{t = 1}^{R}\sum_{r = 1}^{Q_1} \E[G_{H}(\ol \Wb^{t,r-1}, \Hb^{t, r-1})]  +\sum_{t = 1}^{R}\sum_{r =  Q_1 + 1}^{Q^t}\E[G_{W}(\ol \Wb^{t,r-1}, \Hb^{t, r - 1})] \Big]\notag \\
\leq &  \frac{2 (\alpha_1^2 \ol L_H^2 + 1)}{T} \Big(  \alpha_2  \ol L_W \big(F(\ol\Wb^{1, 0}, \Hb^{1, 0}) - \underline{F}\big) \notag\\
& + \frac{16  mk G^2  \ln(1.25 /\delta)   \sum_{t = 1}^{R} (Q_2^t)^3 }{ \alpha_2  \epsilon^{2}}  +    \frac{  \ol L_{W}  \phi^2 \sum_{t = 1}^{R} Q_2^t }{2 \alpha_2 K b \underline{L}_W   }   +  \zeta^2 \Big(\frac{\sum_{t = 1}^{R} Q_2^t}{K}     +   \frac{4 N  \sum_{t = 1}^{R} C_1^t}{ \alpha_2^2 K^2 } \Big) \Big), \label{eqn:model_rate}
\end{align}
\end{small}where
\begin{align}
C_1^t   = Q_2^t(Q_2^t - 1)(2Q_2^t - 1). \label{eqn C1s}
\end{align}
\end{Theorem}
\textit{Proof:}  See Appendix \ref{proof of fedmavg}. \hfill $\blacksquare$

Theorem~\ref{thm: model_avg} provides an upper bound of the  average total local SGDs over $R$ communication rounds; the smaller its value, the higher  convergence rate and the smaller of the cost function in \eqref{eqn: Fed_cluster_prob}  achieved by Algorithm \ref{alg: model_avg}. Based on Theorem~\ref{thm: model_avg}, we have the following   remarks.
\begin{Remark} \label{remark:remark_convergence}
(Convergence rate analysis)
Since $Q_2^t=\lfloor\frac{\widehat{Q}}{t} \rfloor+1$,   we have $\sum_{t=1}^R C_1^t $,  $\sum_{t=1}^R  Q_2^t $ and $\sum_{t=1}^R (Q_2^t)^3$ all in  $ \mathcal{O} (R)$.
According to \eqref{eqn:model_rate},   by setting   $Q_1  = \mathcal{O}(\sqrt{R})$,   the proposed algorithm  converges at a rate of $\mathcal{O}(1/\sqrt{R})$. Furthermore, substituting $T=RQ_1+ \sum_{t = 1}^{R} Q_2^t$ into the bound on the right-hand side of \eqref{eqn:model_rate}, one can infer that the bound decreases with $Q_1$ rather than $Q_2$ due to $Q_2^t \to 1$ as $t$ increases,  implying faster convergence rate for larger $Q_1$ on one hand, and the required DP noise variance  $\sigma_{i,t}^2$ given by (29) is insensitive to $Q_2^t$ on the other hand.

\end{Remark}

\begin{Remark} \label{remark:remark_impact_of_DP}
({Impact of DP}) The larger  value of $\epsilon$ (or $\bar{\epsilon}$), the smaller the upper bound in \eqref{eqn:model_rate}, implying that the better  learning performance (convergence rate and the loss function $F$) and the weaker   required privacy protection level, namely a privacy-utility tradeoff.
\end{Remark}

\vspace{-0.5cm}
\begin{Remark} \label{remark:remark_impact_of_localSGD}
({Impact of non-i.i.d. data and PCP})
The smaller the value of $\zeta$ or the larger the value of $K$, the smaller the upper bound in \eqref{eqn:model_rate}, implying   the smaller degree of non-i.i.d. data or the more clients in PCP, and the better  learning performance (faster convergence rate and smaller loss function $F$).
\end{Remark}
\vspace{-0.3cm}

\begin{Remark} \label{remark:complexity_comparison}
(Complexity comparison with existing federated clustering methods)
Suppose that all clients participate the model training ($N$ clients),  the complexity of federated $k$-means (FKM) for each local iteration at the client side is  $\mathcal{O}\left(m n k N   + m n N (\log N)^2 \right)$, and the complexity at the  PS side is $\mathcal{O}\left(m  n k \right)$~\cite{li2022secure}.
It can be verified that the per-iteration complexity for the proposed DP-FedC algorithm is $\mathcal{O}\left( (mN + n)k^2 + m n k) \right)$ at the client side, and a complexity order of $\mathcal{O}(mkN)$ at the PS side (shown in Appendix~\ref{sec:proof Lemma6}).   As a result, the complexity of the proposed DP-FedC algorithm is smaller than that of FKM since  $ k  < N  \ll n$  is true in general. However, the DP-FedC and the FZKM~\cite{stallmann2022towards} have comparable complexity at both client side and the PS side, simply because they have similar computing procedure, in spite of no complexity analysis reported in \cite{stallmann2022towards}.
\end{Remark}


\section{Experiment Results}\label{sec: simulation}
In this section, in terms of the cost function (i.e., the objective value)   in \eqref{eqn: Fed_cluster_prob} and clustering accuracy, some experimental results are presented to evaluate the performance of the proposed DP-FedC algorithm (Algorithm \ref{alg: model_avg}) including comparison with some state-of-the-art FedC algorithms. The experiment is performed using two real datasets and each obtained result is the average over 5 independent runs with the same randomly generated initial feasible points for all the algorithms under test.

\vspace{-0.2cm}
\subsection{Experiment setup}

{\bf Datasets:} The two real data sets used in the experiment are TCGA \cite{TCGA_CGCD} and MNIST datasets.
Specifically,  TCGA dataset was obtained from the Cancer Genome Atlas database which contains the gene expression data of 5,314 cancer samples belonging to 20 cancer types.
Each data sample in TCGA dataset is a real column vector containing the top-ranked 5000 features selected through  Pearson's Chi-Squares Test~\cite{wang2022federated}.
The MNIST database contains 60,000 training images of 10 handwritten digits and 10,000 test ones. We randomly select 10,000 images from the 60,000 training images as the   dataset in our experiment, where each data sample is a real column vector containing 784 features.   These two datasets are representative, i.e., one (the other) with large (small) data size but small (large) feature size, also implying challenging unsupervised clustering for both datasets in our experiment.

In the experiment, we distribute the samples of each dataset to $N=100$ clients in the following two ways:
\begin{enumerate}[(i)]
\item \textbf{IID case}: We follow the data partition method in \cite{DKCoreset_2013} to obtain balanced and i.i.d. distributed data for the two datasets. To be specific, the i.i.d. distributed data are generated by randomly assigning the data samples to all clients.
\item \textbf{non-IID case}: For the TCGA dataset, we apply the $k$-means algorithm to cluster the dataset into $100$ clusters, and the data samples belonging to the same  cluster is assigned to one client.
    For the MNIST dataset, we follow the partition method in \cite{li2019convergence} to obtain distributed data such that each client's dataset only contains two digits, thus yielding a highly unbalanced and non-i.i.d. dataset.
\end{enumerate}

{\bf Parameter setting:} In problem~\eqref{eqn: Fed_cluster_prob}, if not mentioned specifically, we set the parameters as follows: $k=10$ for MNIST and $k=20$ for TCGA, $\mu_{w}=0$, $\rho=10^{-7} \times \frac{\|\Xb\|_F^2}{N}$ and $\mu_{h} = 10^{-10} \times \frac{\|\Xb\|_F^2}{N}$. As for the parameters in Algorithm \ref{alg: model_avg},  the step size $\gamma_i^{t} =\frac{1}{2} L_{H_i}^t$ where $L_{H_i}^t$ is estimated as $\lambda_{\max}((\Wb_i^{t,0})^\top \Wb_i^{t,0})$. Analogously, the step size $\eta^t =5L_{W}^t$ where $L_{W}^t$ is estimated as $\lambda_{\max}(\Hb^{t, Q_1}(\Hb^{t, Q_1})^\top)$.  In all experiments, we assume all clients have the same privacy protection level (i.e., $\epsilon_i=\epsilon$, for all $i$) and the same total privacy loss budget  (i.e., $\bar{\epsilon}_i=\bar{\epsilon}$ for all $i$). Then, given the total privacy loss $\bar{\epsilon}$,  the privacy protection level $\epsilon$ at each communication round is obtained by Theorem~\ref{Thm:total_budget} for $R=100$ and $\delta=10^{-4}$. The mini-batch dataset size $b$ is set to   $50$.
Other parameters are empirically chosen to our best. All the algorithms under test run  until $R = 100$ is reached. Then the clustering accuracy is calculated as the ratio of the number of correct classifications (no. of columns of all the estimated ${\bf H}_i,i\in[N]$, i.e., their maximum column  entries falling  in the correct cluster) to the total number of data (i.e., $ n$).

\vspace{-0.1cm}
\subsection{Impact of DP}\label{subsec: conv of FedAM}
\begin{figure*}[t!]
\begin{minipage}[b]{0.24\linewidth}
\centering
\includegraphics[scale=0.30]{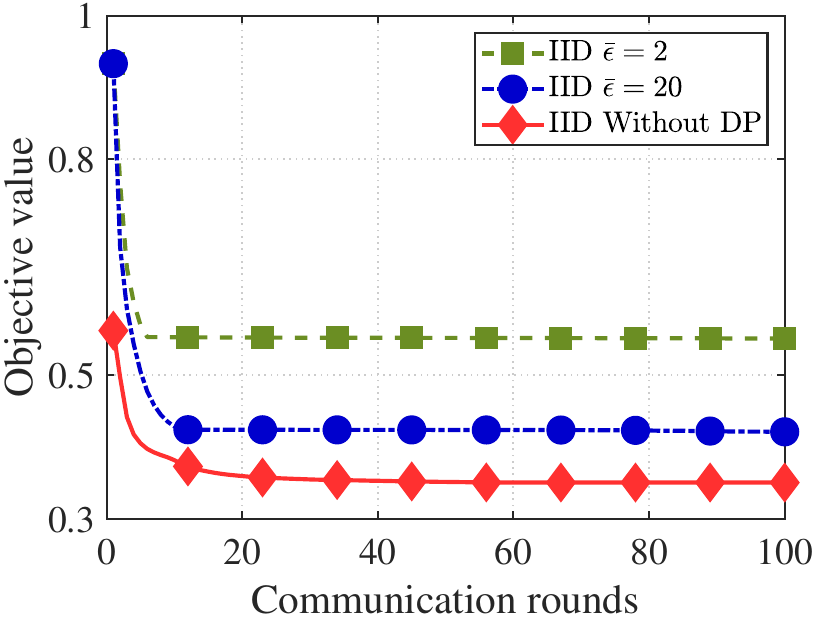}
\centerline{\scriptsize{(a)  MNIST, {\bf IID}  }}\medskip
\end{minipage}
\hfill
\begin{minipage}[b]{0.24\linewidth}
\centering
\includegraphics[scale=0.30]{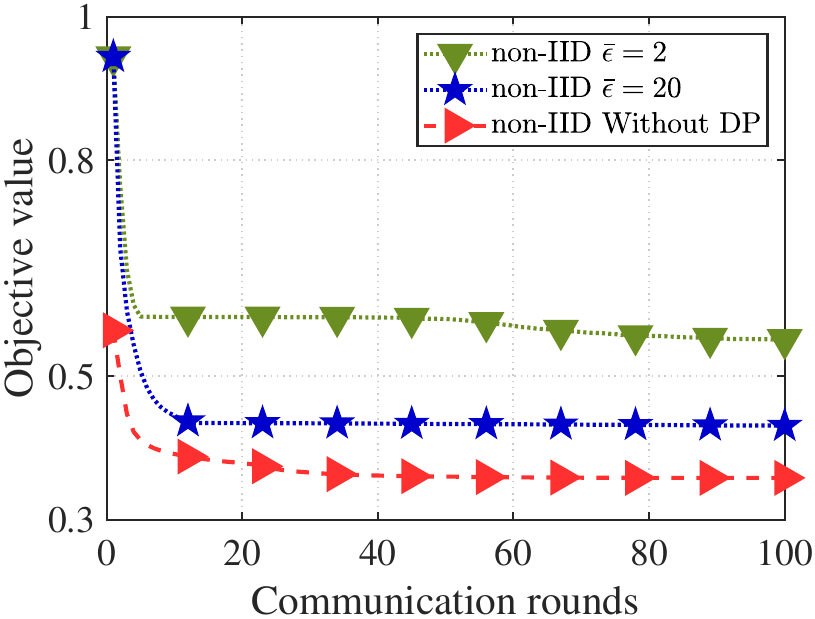}
\centerline{\scriptsize{(b)  MNIST, {\bf non-IID} }}\medskip
\end{minipage}
\begin{minipage}[b]{0.24\linewidth}
\centering
\includegraphics[scale=0.30]{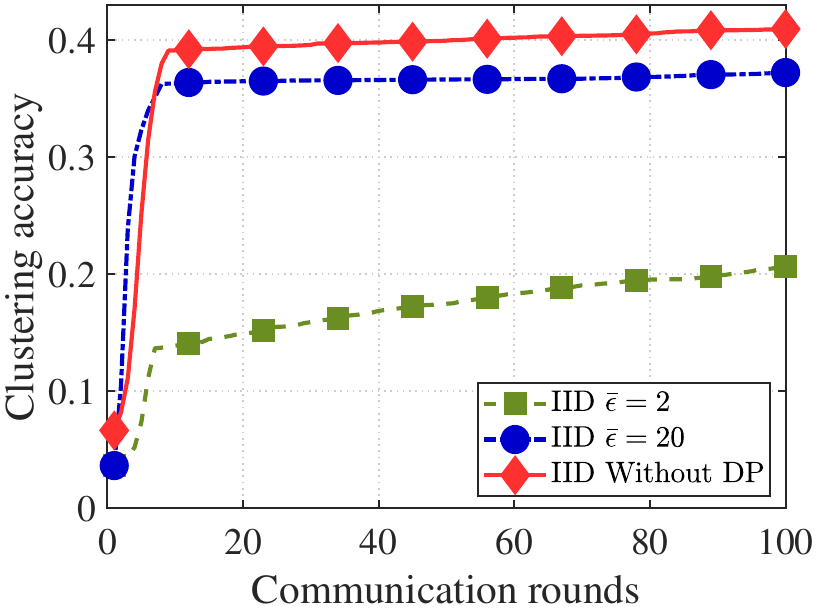}
\centerline{\scriptsize{(c)   MNIST, {\bf IID}  }}\medskip
\end{minipage}
\hfill
\begin{minipage}[b]{0.24\linewidth}
\centering
\includegraphics[scale=0.30]{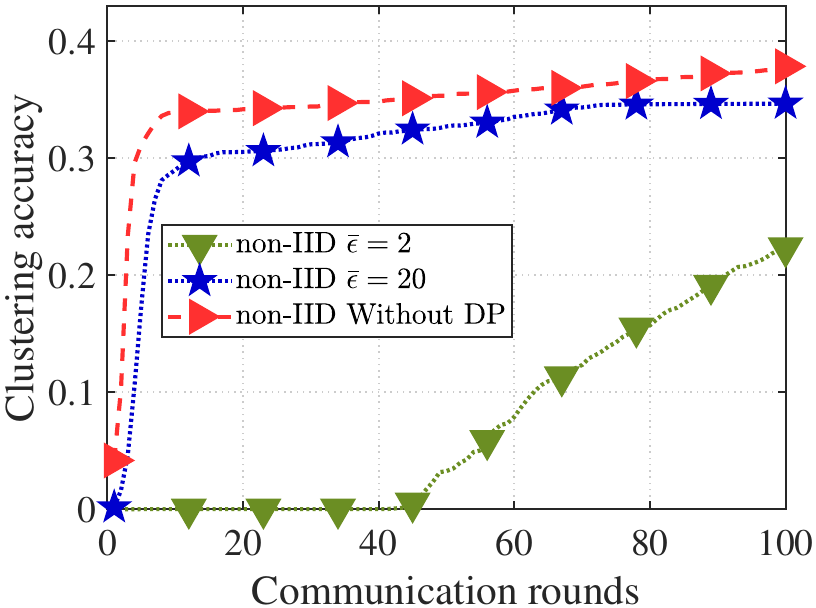}
\centerline{\scriptsize{(d)    MNIST, {\bf non-IID} }}\medskip
\end{minipage}
\begin{minipage}[b]{0.24\linewidth}
\centering
\includegraphics[scale=0.30]{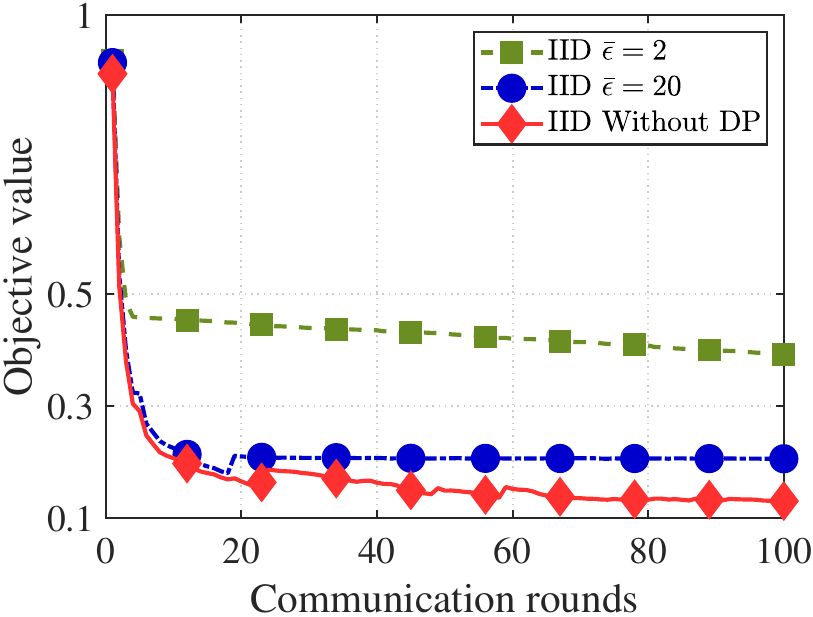}
\centerline{\scriptsize{(e)  TCGA, {\bf IID}  }}\medskip
\end{minipage}
\hfill
\begin{minipage}[b]{0.24\linewidth}
\centering
\includegraphics[scale=0.30]{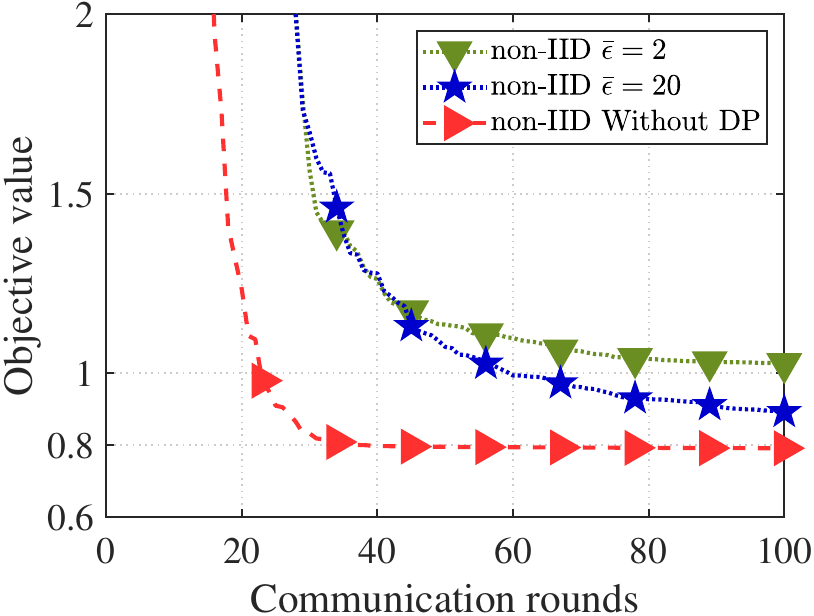}
\centerline{\scriptsize{(f)  TCGA, {\bf non-IID}  }}\medskip
\end{minipage}
\begin{minipage}[b]{0.24\linewidth}
\centering
\includegraphics[scale=0.30]{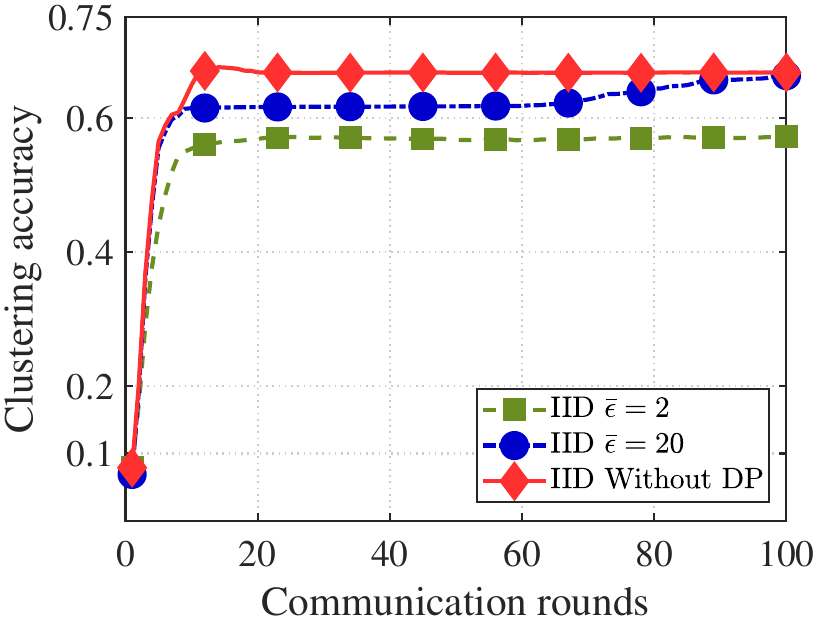}
\centerline{\scriptsize{(g)   TCGA,  {\bf IID}  }}\medskip
\end{minipage}
\hfill
\begin{minipage}[b]{0.24\linewidth}
\centering
\includegraphics[scale=0.30]{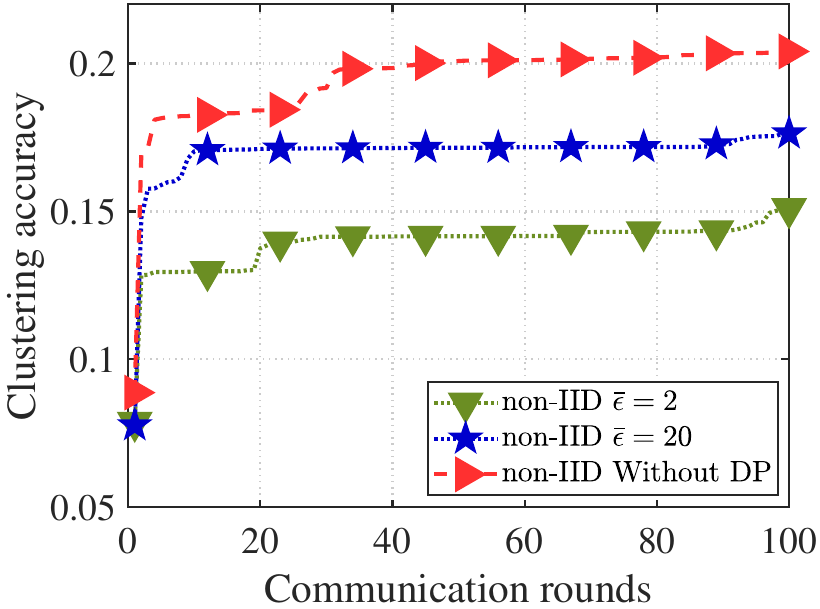}
\centerline{\scriptsize{(h)   TCGA, {\bf non-IID} }}\medskip
\end{minipage}
\caption{Objective   value  and clustering accuracy versus communication rounds of the proposed DP-FedC algorithm for {\bf IID} case and {\bf non-IID} case, where (a)-(d), and (e)-(f), are obtained using MNIST dataset and TCGA dataset, respectively, for  the cases of without DP, and $\bar\epsilon \in \{2,20\}$.}
\vspace{-0.15cm}
\label{epsilon_accuracy}
\end{figure*}

Figure  \ref{epsilon_accuracy} depicts the objective value for simplicity (i.e., the value  of $F({\bf W},{\bf H})$ in \eqref{eqn: Fed_cluster_prob})   and the clustering accuracy   versus communication round  with different values of $\bar{\epsilon}$ for both \textbf{IID case} and \textbf{non-IID case}, where $K=30$, $Q_1=10$, and $Q_2^t=\lfloor\frac{10}{t} \rfloor+1$.
Some observations from Figs. \ref{epsilon_accuracy}(a)-(d), are as follows:
\begin{itemize}
  \item [(i)]  The larger the value of $\bar{\epsilon}$ where the results without DP conceptually corresponds to $ \bar{\epsilon}\rightarrow \infty $, the smaller the objective value and the higher the clustering accuracy and convergence rate  for both \textbf{IID case} case and \textbf{non-IID case};
  \item [(ii)] The   objective value is smaller and the clustering accuracy is higher for the \textbf{IID case} than for \textbf{non-IID case}, and the performance gap between the two cases seems more appreciable in  clustering accuracy.
\end{itemize}
The above two observations also apply to Figs. \ref{epsilon_accuracy}(e)-(h). Moreover, the impact of non-i.i.d. data is more serious on the TCGA dataset. These results are consistent with Remark~\ref{remark:remark_impact_of_DP} and Remark~\ref{remark:remark_impact_of_localSGD}, so a proper choice of $\bar{\epsilon}$ value is needed to achieve a good privacy-utility tradeoff.

\vspace{-0.2cm}
\subsection{Impact of the number of participated clients ($K$)}\label{subsec: conv of fedcgds}
\begin{figure*}[t!]
\begin{minipage}[b]{0.24\linewidth}
\centering
\includegraphics[scale=0.30]{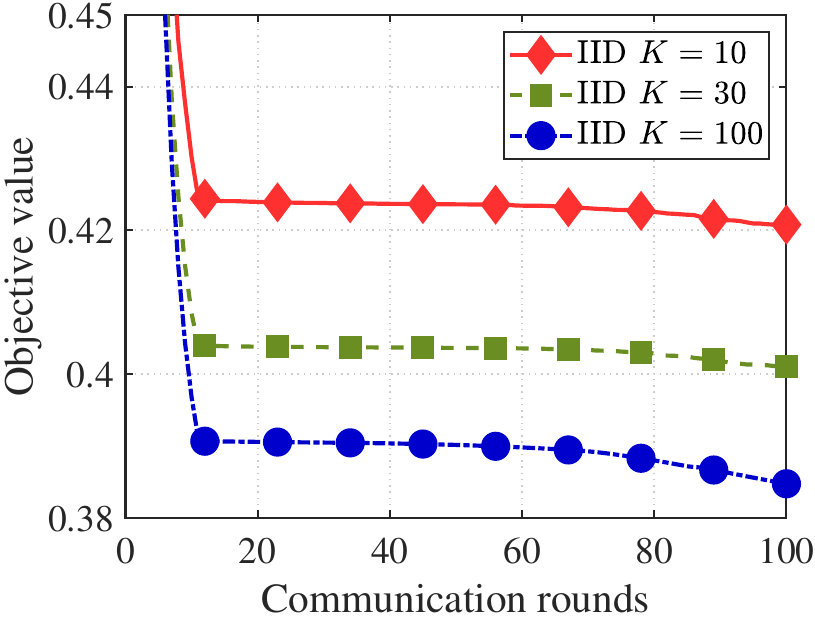}
\centerline{\scriptsize{(a)  MNIST, {\bf IID}  }}\medskip
\end{minipage}
\hfill
\begin{minipage}[b]{0.24\linewidth}
\centering
\includegraphics[scale=0.30]{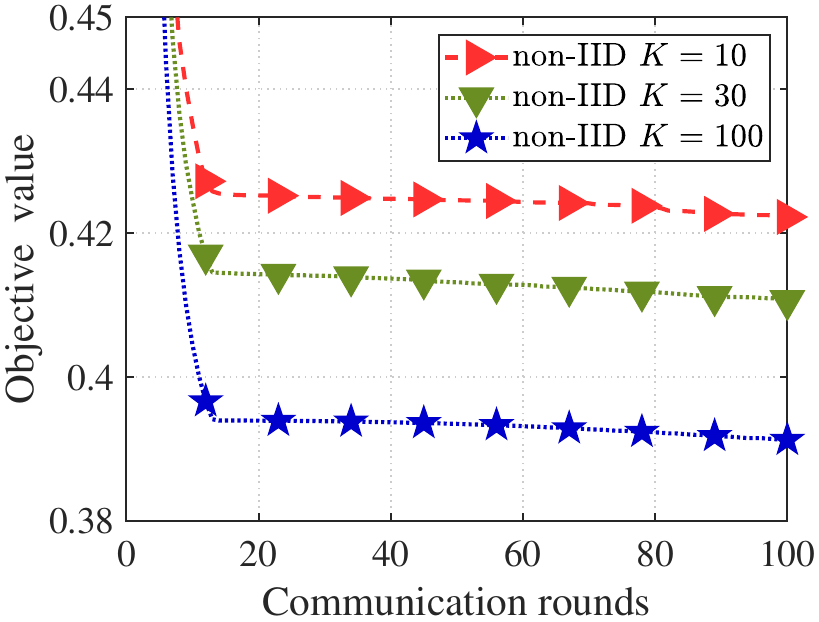}
\centerline{\scriptsize{(b) MNIST, {\bf non-IID} }}\medskip
\end{minipage}
\begin{minipage}[b]{0.24\linewidth}
\centering
\includegraphics[scale=0.30]{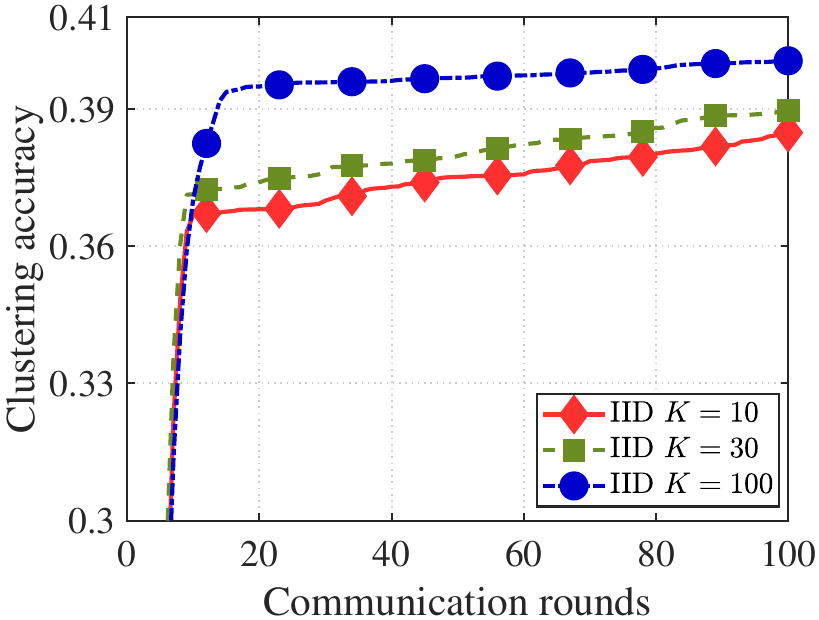}
\centerline{\scriptsize{(c) MNIST, {\bf IID}  }}\medskip
\end{minipage}
\hfill
\begin{minipage}[b]{0.24\linewidth}
\centering
\includegraphics[scale=0.30]{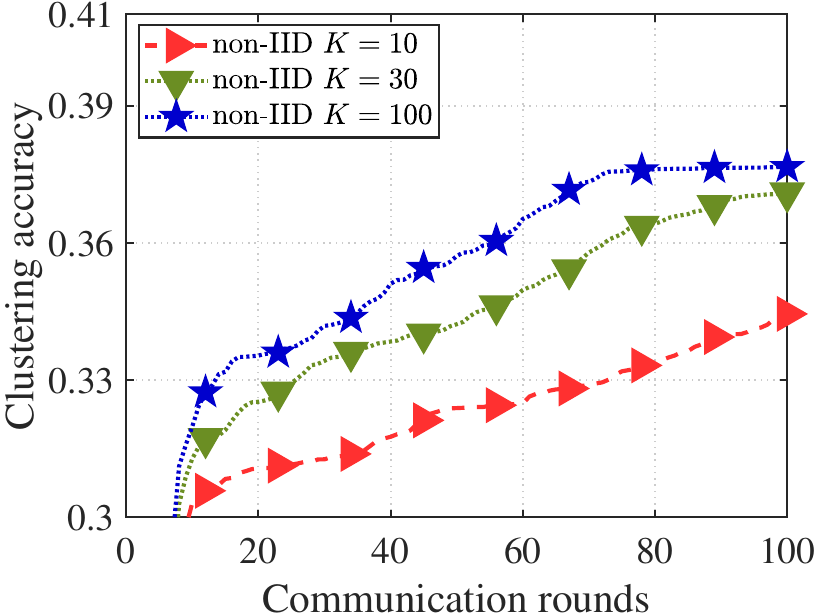}
\centerline{\scriptsize{(d)    MNIST, {\bf non-IID} }}\medskip
\end{minipage}
\begin{minipage}[b]{0.24\linewidth}
\centering
\includegraphics[scale=0.30]{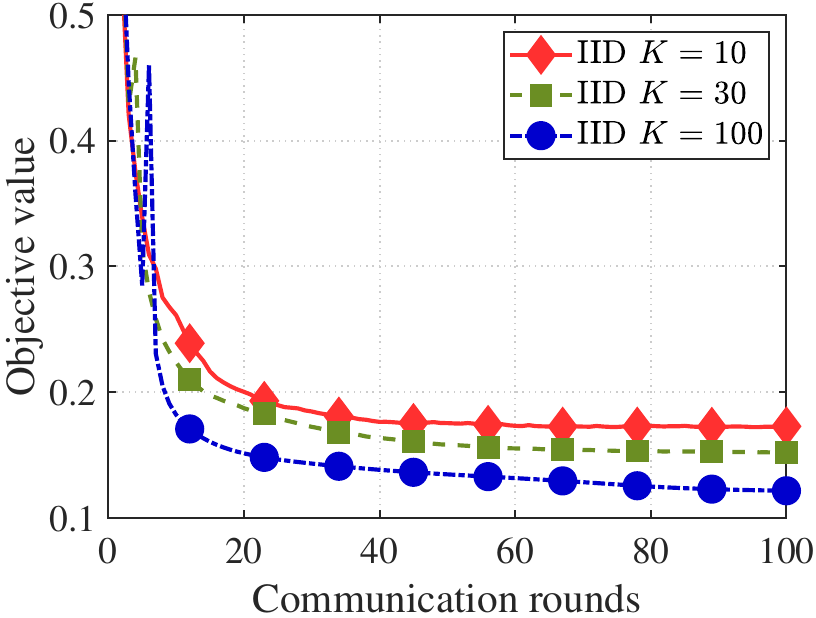}
\centerline{\scriptsize{(e)  TCGA, {\bf IID}  }}\medskip
\end{minipage}
\hfill
\begin{minipage}[b]{0.24\linewidth}
\centering
\includegraphics[scale=0.30]{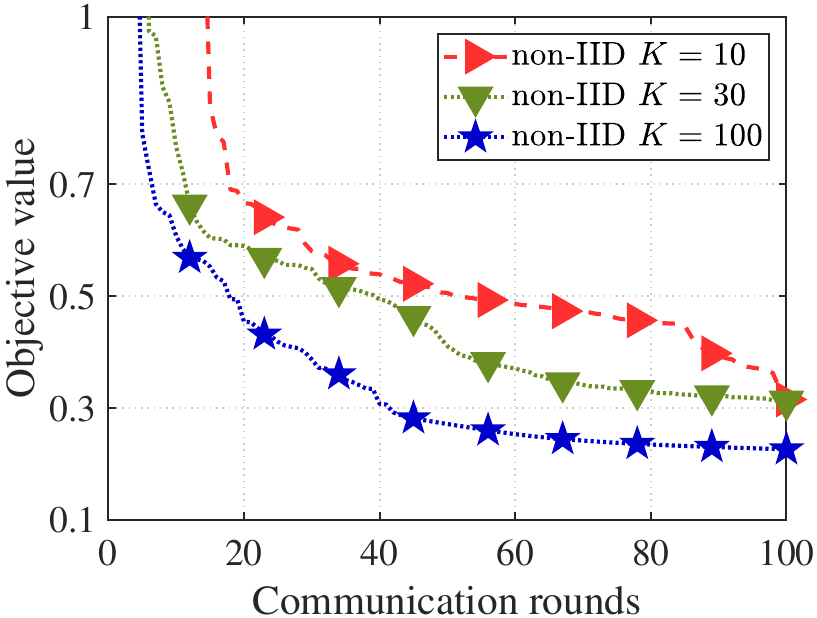}
\centerline{\scriptsize{(f)  TCGA, {\bf non-IID}  }}\medskip
\end{minipage}
\begin{minipage}[b]{0.24\linewidth}
\centering
\includegraphics[scale=0.30]{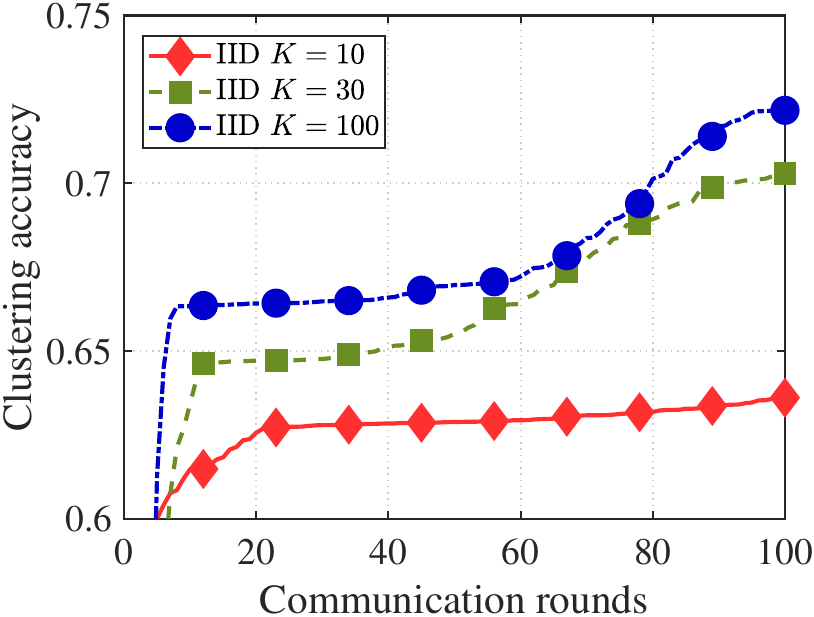}
\centerline{\scriptsize{(g)   TCGA,  {\bf IID}  }}\medskip
\end{minipage}
\hfill
\begin{minipage}[b]{0.24\linewidth}
\centering
\includegraphics[scale=0.30]{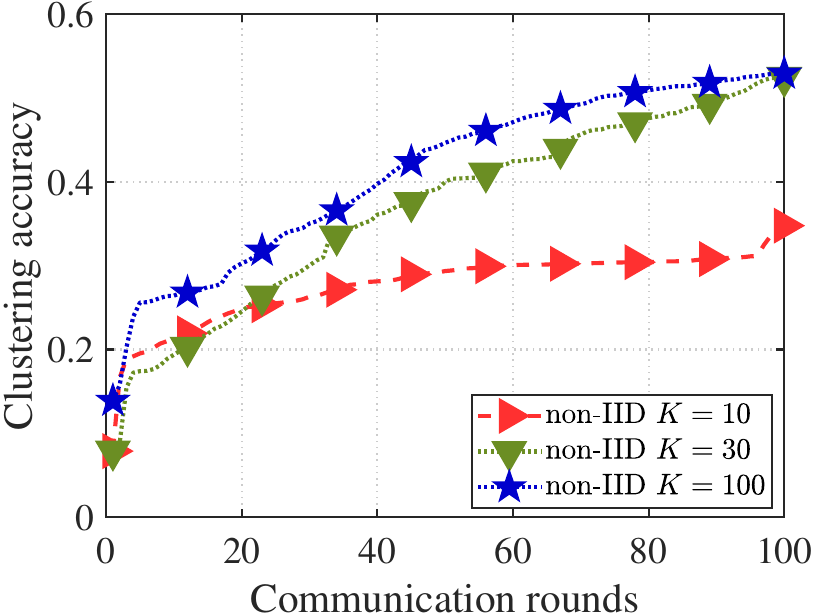}
\centerline{\scriptsize{(h)   TCGA, {\bf non-IID} }}\medskip
\end{minipage}
\caption{Objective   value    and clustering accuracy versus communication rounds of the proposed DP-FedC algorithm for {\bf IID} case and {\bf non-IID} case, where (a)-(d), and (e)-(f), are obtained using MNIST dataset and TCGA dataset, respectively, for $\bar{\epsilon}=20$ and $K\in\{10,30,100\}.$}
\vspace{-0.15cm}
\label{client_accuracy}
\end{figure*}

\vspace{-0.0cm}
Figure~\ref{client_accuracy} depicts the convergence performance of DP-FedC versus communication rounds under different values of $K$ with $\bar{\epsilon}=20$,   $Q_1=10$, and $Q_2^t=\lfloor\frac{10}{t} \rfloor+1$.
It can be seen from Figs. \ref{client_accuracy}(a), 4(b), 4(e) and 4(f), that the objective value is smaller together with faster convergence rate either for larger $K$ or for the {\bf IID} case. This is also true for the clustering accuracy, though the convergence rate on TCGA for the {\bf  IID} case is only slightly better than for the {\bf non-IID} case. These results are also consistent with Remark~\ref{remark:remark_impact_of_localSGD}.

\subsection{Comparison with existing distributed clustering methods}
We here compare the proposed DP-FedC algorithm with four benchmark algorithms in terms of clustering performance.  These algorithms include federated $k$-means (FKM)~\cite{li2022secure}, federated fuzzy $k$-means (FZKM)~\cite{stallmann2022towards},  distributed $k$-means++ (DK++)~\cite{Kmeans_2012},  distributed $k$-median (DKM)~\cite{DKCoreset_2013}. The first two are   state-of-the-art federated clustering algorithms while the latter two are traditional  distributed clustering methods. As mentioned previously, they were   basically developed by extending  the $k$-means algorithm and its variants.
We add the artificial noise to DP noise that guarantees the $(\epsilon,\delta)$-DP  at each communication round in the implementation of the above four existing algorithms in our experiment. Then we apply the proposed algorithm to process the given dataset with parameters $K=30$, $Q_1=10$, $Q_2^t=5$, $\bar{\epsilon} = 20$ under the i.i.d. data case. However, the parameters used for the other four algorithms are taken from the associated references together with $K=30$, $\bar{\epsilon} = 20$.

The obtained experimental results (for the clustering accuracy) are  listed in Table \ref{tab:table1}. It can be seen from this table that   the clustering accuracy performances of all the algorithms under test   for the case of without DP noise are better than with DP noise used. The performance gap between the two cases for our DP-FedC algorithm is much smaller than for the other algorithms, implying that the proposed algorithm is more robust again DP noise thanks to the privacy amplification strategy applied.

\begin{table*}[t!]
\centering
\caption{Performance comparison  of five algorithms   in terms of clustering accuracy (\%).}
\label{tab:table1}
\setlength{\tabcolsep}{6.8mm}
\begin{tabular}{|c|c|c|c|c|}
\hline    \rowcolor{gray!20}
\diagbox{Method}{Dataset}       &       \begin{tabular}[c]{@{}c@{}}TCGA\\  (without DP) \end{tabular}         &      \begin{tabular}[c]{@{}c@{}}MNIST\\  (without DP) \end{tabular}
&     \begin{tabular}[c]{@{}c@{}}TCGA\\  ($\bar{\epsilon}=20$) \end{tabular}      &    \begin{tabular}[c]{@{}c@{}}MNIST\\  ($\bar{\epsilon}=20$) \end{tabular}  \\  \hline\hline
DK++~\cite{Kmeans_2012}     &     65.4    &      42.6      &     50.1    &   26.8    \\ \cline{1-5}
DKM~\cite{DKCoreset_2013}        &    38.7     &   43.3          &        31.0    &  26.1   \\ \cline{1-5}
\tabincell{c}{FKM}~\cite{li2022secure}        &     70.2           &      43.2      &     58.4   &  31.8     \\ \cline{1-5}
\tabincell{c}{FZKM}~\cite{stallmann2022towards}       &     72.8             &    47.1      &   66.9     &    36.4     \\ \cline{1-5}	
\tabincell{c}{{\bf DP-FedC} }      & \bf{76.7}    & \textbf{50.5 } & \textbf{72.2}  & \bf{43.1}    \\ \cline{1-5}			
\hline
\end{tabular}
\vspace{-0.45cm}
\end{table*}

\section{Conclusion}\label{sec: conclusion}
We have presented a novel FedC algorithm called DP-FedC (Algorithm~\ref{alg: model_avg}), which is based on the traditional clustering algorithm $k$-means and operates according to the computation-aggregation protocol. Specifically, the proposed DP-FedC employs DP-based privacy protection, along with the policies of PCP and multiple local SGD updating steps implemented in the algorithm design. Various characteristics and insights of Algorithm~\ref{alg: model_avg} were discovered through theoretical analyses, including the impact of system parameters on privacy amplification, convergence rate, and the impact of data heterogeneity (e.g., non-i.i.d. data) on learning performance. These analytical results can serve as valuable guidelines for practical FL algorithm design, especially when considering the preferred tradeoff between learning performance and the required level of privacy protection.  Finally, we provided experimental results on two real datasets to demonstrate the efficacy of the proposed method, along with its superior performance over state-of-the-art FedC algorithms, and its consistency with all the presented analytical results.

\numberwithin{equation}{section}
\appendices

\section{Proof of Theorem \ref{thm: privacy amplicfication via sampling} }\label{appendix:proof Theorem1}
The proof mainly follows the work~\cite{balle2018privacy} by considering  both data sampling  with replacement case and that  without replacement case.

Suppose that data subsampling mechanism yields $(\epsilon^{\prime},   \delta^{\prime})$-DP, when  $\epsilon \leq 1$ and data  are uniformly sampled with replacement,   data  subsampling mechanism guarantees $(\ln(1 + q(\exp(\epsilon)-1), q\delta)$-DP~\cite{balle2018privacy},  then, we have $\delta^{\prime} = q \delta$ and
\begin{align}
\epsilon^{\prime}  =  \ln(1+q(\exp(\epsilon)-1)) \overset{(a)}{\leq} q(\exp(\epsilon) -1) \overset{(b)}{\leq} 2 q \epsilon, \label{eqn:A_1}
\end{align}where $(a)$ and $(b)$ hold because   $\ln(1+x) \leq x$ and $\exp(x)-1 \leq 2x$ when $0 < x \leq 1$~\cite{ChiBook2017}.

When data are uniformly sampled without replacement, we still have $\delta^{\prime} = q \delta$ and $\epsilon^{\prime}$ becomes~\cite{balle2018privacy},
\begin{align}
\epsilon^{\prime} &= \ln\big(1+(1-(1- \frac{1}{n}))^{b} (\exp(\epsilon)-1)\big) \notag\\
& \overset{(a)}{\leq} \ln \big(1+ q (\exp(\epsilon)-1) \big)  \overset{(b)}{\leq}  2 q \epsilon, \label{eqn:A_2}
\end{align}where $(b)$ follows because of \eqref{eqn:A_1}, and $(a)$ holds since
\begin{align}\label{eqn:A_3}
& (1-(1- \frac{1}{n}))^{b} \leq  \frac{b}{n} = q.
\end{align}
By combining \eqref{eqn:A_2} and \eqref{eqn:A_3}, we   obtain $\epsilon^{\prime} \leq 2 q \epsilon$ for data sampling without replacement.

Then, when $q \geq 1/2$ (i.e., $2q\epsilon > \epsilon$), there is no privacy amplification. In this case, we have
\begin{align}\label{eqn:A_4}
\epsilon^{\prime}= \epsilon.
\end{align}
Therefore, by combining \eqref{eqn:A_1}, \eqref{eqn:A_2} and \eqref{eqn:A_4}, we have $\epsilon^{\prime} = \min (2q \epsilon,\epsilon)$. Thus, we complete the proof.  $\hfill\blacksquare$

\section{Proof of Lemma \ref{lemma:sensitivity} }\label{proof Lemma_sensitivity}
Assume  $\mathcal{D}_{i}$ and $\mathcal{D}_{i}^{\prime}$ are the neighboring datasets that differ in only one data sample.  Without loss of generality, let $u_{i}$ be  the unique different element between $\mathcal{D}_{i}$ and $\mathcal{D}_{i}^{\prime}$, i.e., $ \mathcal{D}_{i}^{\prime} \cup \{u_{i}\} = \mathcal{D}_i \cup \{u_{i}\}$.
For clarity of the following proof, let us make the following notational correspondences:
${\Wb}_i^{t,r} \leftrightarrow \Wb_{\Dc_i}^{t,r}$,
${\Hb}_i^{t,r} \leftrightarrow {\Hb}_{\Dc_i}^{t,r}$, and $\mathcal{B}_i^{t,r} \leftrightarrow \mathcal{B}_{\Dc_i}^{t,r}$.
Then, for any $r \in [Q^t]\setminus [Q_1]$, the $\ell_{2}$-sensitivity~\cite{dwork2014algorithmic} of $\Wb_{i}^t$  is calculated by
\begin{small}
\begin{align}\label{eqn:sensitivity_y}
s_{i}^{t}  =&  \max _{\Dc_{i}, \mathcal{D}_{i}^{\prime}} \big\|\Wb_{\mathcal{D}_{i}}^{t} -  \Wb_{\mathcal{D}_{i}^{\prime}}^{t}\big\|  \notag\\
= & \max _{\mathcal{D}_{i}, \mathcal{D}_{i}^{\prime}} \Big\| \sum_{r=Q_1+1}^{Q^t} \Wb_{ \mathcal{D}_{i}}^{t, r-1} - \frac{\nabla_{W}F_i(\Wb_{ \mathcal{D}_{i}}^{t, r-1}, \Hb_{ \mathcal{D}_{i}}^{t, r-1}; \mathcal{B}_{ \mathcal{D}_{i}}^{t,r})}{ \eta^t} - \sum_{r=Q_1+1}^{Q^t}\big( \Wb_{ \mathcal{D}_{i}^{\prime}}^{t, r-1} - \frac{\nabla_{W}F_i(\Wb_{ \mathcal{D}_{i}^{\prime}}^{t, r-1}, \Hb_{ \mathcal{D}_{i}^{\prime}}^{t, r-1}; \mathcal{B}_{ \mathcal{D}_{i}^{\prime}}^{t,r})}{ \eta^t} \big) \Big\|  \notag\\
= & \max _{\mathcal{D}_{i}, \mathcal{D}_{i}^{\prime}} \Big\|   \Big( \Wb_{ \mathcal{D}_{i}}^{t, Q_1} -  \frac{\nabla_{W}F_i(\Wb_{ \mathcal{D}_{i}}^{t, Q_1}, \Hb_{ \mathcal{D}_{i}}^{t, Q_1}; \mathcal{B}_{ \mathcal{D}_{i}}^{t,Q_1+1})}{ \eta^t} - \cdots -\frac{\nabla_{W}F_i(\Wb_{ \mathcal{D}_{i}}^{t, Q^t-1}, \Hb_{  \mathcal{D}_{i}}^{t, Q^t-1}; \mathcal{B}_{ \mathcal{D}_{i}}^{t,Q^t})}{ \eta^t} \Big) \notag\\
& -  \Big( \Wb_{ \mathcal{D}_{i}^{\prime}}^{t, Q_1} -  \frac{\nabla_{W}F_i(\Wb_{ \mathcal{D}_{i}^{\prime}}^{t, Q_1}, \Hb_{ \mathcal{D}_{i}^{\prime}}^{t, Q_1}; \mathcal{B}_{ \mathcal{D}_{i}^{\prime}}^{t,Q_1+1})}{ \eta^t} - \cdots -\frac{\nabla_{W}F_i(\Wb_{ \mathcal{D}_{i}^{\prime}}^{t,  Q^t-1}, \Hb_{ \mathcal{D}_{i}^{\prime}}^{t,  Q^t-1}; \mathcal{B}_{ \mathcal{D}_{i}^{\prime}}^{t, Q^t})}{ \eta^t} \Big) \Big\|  \notag\\
\overset{(a)}{\leq}& \frac{2 G Q_2^t }{ \eta^t},
\end{align}
\end{small}where $(a)$ holds because of Assumption \ref{Ass: bounded_gradient}, and  $\Wb_{ \mathcal{D}_{i}}^{t, Q_1}=\Wb_{ \mathcal{D}_{i}^{\prime}}^{t, Q_1}$ always holds.   $\hfill\blacksquare$

\section{Proof of Theorem \ref{thm: model_avg}}\label{proof of fedmavg}
According to \eqref{eqn: q_1} and \eqref{eqn: FedAM update of W2 0}, we have
\begin{small}
\begin{align}\label{eqn: q_1_appendix}
\!\!\!\ol \Wb^{t, r} =&   \ol \Wb^{t, r-1} - \frac{1}{K \eta^t}  \sum_{i \in \Sc^t} \nabla_{W}F_i(\Wb_i^{t, r-1}, \Hb_i^{t, r-1};\mathcal{B}_i^{t,r}).
\end{align}
\end{small}

\noindent  ({\bf Objective Descent w.r.t. $\Hb$})
According to \cite[Lemma 3.2]{BOLTE2014PROXIMAL} and setting $\gamma_i^{t}= \alpha_1  L_H^t/2 \leq \alpha_1 \ol L_H/2$ where  $\alpha_1 >1$,  we have
\begin{small}\begin{align}
F_i(\ol\Wb^{t,r}, \Hb_i^{t, r}) - F_i(\ol\Wb^{t,r-1}, \Hb_i^{t, r-1}) \leq -  \frac{\alpha_1 - 1}{2}\ol L_H    \|\Hb_i^{t, r-1} - \Hb_i^{t, r}\|_F^2, \forall r \in [Q_1]. \label{thm1: descent_H}
\end{align}\end{small}
Taking expectation over two sides of \eqref{thm1: descent_H} and then summing up  from $r = 1$ to $Q_1$   yields

\vspace{-0.03cm}
\begin{small}\begin{align}
\E\big[F_i(\ol\Wb^{t,Q_1}, \Hb_i^{t, Q_1})\big] - \E\big[F_i(\ol\Wb^{t,0}, \Hb_i^{t, 0})\big] \leq - \frac{\alpha_1 - 1}{2}\ol L_H \sum_{r = 1}^{Q_1} \E\big[\|\Hb_i^{t, r-1} - \Hb_i^{t, r}\|_F^2\big], \forall r \in [Q_1]. \label{thm1: obj_descent_Hp}
\end{align}\end{small}
\vspace{-0.03cm}

By taking the summation  over two sides of \eqref{thm1: obj_descent_Hp} from $i=1$ to $N$, the objective function $F$ descends with local updates of $\Hb$ is given by

\vspace{-0.3cm}
\begin{small}\begin{align}
\E[F(\ol\Wb^{t,Q_1}, \Hb^{t, Q_1})] - \E[F(\ol\Wb^{t,0}, \Hb^{t, 0})] \leq -\frac{\alpha_1 - 1}{2}\ol L_H \sum_{r = 1}^{Q_1}\sum_{i = 1}^{N}  \E\big[\|\Hb_i^{t, r - 1} - \Hb_i^{t, r}\|_F^2\big], \forall r \in [Q_1]. \label{thm1: descent_H2}
\end{align}\end{small}
\vspace{-0.1cm}

\noindent ({\bf Objective Descent w.r.t. $\Wb$})
Since $\Hb_i^{t, r} = \Hb_i^{t, r-1}$ (cf. line 14 in Algorithm~\ref{alg: model_avg}) and $\nabla_{W} F(\cdot, \Hb^{t,Q})$ is Lipschitz continuous under Assumption \ref{Ass: bounded_Lipschitz}. Then,  by the descent lemma \cite[Lemma 3.1]{BOLTE2014PROXIMAL}, when $ r \in [Q^t-1]\setminus [Q_1]$,  we have

\vspace{-0.4cm}
\begin{small}
\begin{align}
\E \big[F(\ol\Wb^{t, r},\Hb^{t, r}) \big] \leq&  \E \big[F(\ol\Wb^{t, r-1}, \Hb^{t, r-1}) \big] +  \frac{L_{W}^t}{2} \E \big[ \|\ol\Wb^{t, r} - \ol\Wb^{t, r-1}\|_F^2 \big] \notag \\
&+  \E \big[\langle \nabla_{W} F(\ol\Wb^{t, r-1}, \Hb^{t, r-1}), \ol\Wb^{t, r} - \ol\Wb^{t, r-1} \rangle \big].\label{thm1: lip_W_1}
\end{align}
\end{small}
When $r =Q^t$,  by Algorithm \ref{alg: model_avg}, \eqref{thm1: lip_W_1} becomes,
\begin{small}
\begin{align}
\E \big[F(\ol\Wb^{t, Q^t},\Hb^{t, Q^t}) \big] \leq&  \E \big[F(\ol\Wb^{t, Q^t-1}, \Hb^{t, Q^t-1}) \big] +  \underbrace{ \frac{L_{W}^t}{2} \E \big[ \|\ol\Wb^{t, Q^t} - \ol\Wb^{t, Q^t-1}+\xib^t\|_F^2 \big]}_{\triangleq(\rm S.1)} \notag \\
&+   \E \big[\langle \nabla_{W} F(\ol\Wb^{t, Q^t-1}, \Hb^{t, Q^t-1}), \ol\Wb^{t, Q^t} - \ol\Wb^{t, Q^t-1} \rangle \big],  \label{thm1: lip_W_Qt}
\end{align}
\end{small}where $\xib^t = \frac{1}{K}\sum_{i=i}^{K} \xib_i^t$. The $(\rm S.1)$ can be further bounded by,
\begin{small}
\begin{align}
\rm  (S.1) =& \frac{L_{W}^t}{2} \E\big[ \|\ol \Wb^{t,  Q^t} - \ol \Wb^{t,  Q^t-1} + \xib^t \|_F^2\big] \notag \\
=& \frac{L_{W}^t}{2} \E\big[ \|\ol \Wb^{t,  Q^t-1} - \ol \Wb^{t,  Q^t}  \|_F^2] + \frac{L_{W}^t}{2} \E\big[ \| \xib^t \|_F^2\big] \notag\\
\overset{(a)}{\leq}&  \frac{L_{W}^t}{2} \E\big[ \|\ol \Wb^{t,  Q^t-1} - \ol \Wb^{t,  Q^t}  \|_F^2\big] + \frac{16 mk G^2  \ln(1.25 /\delta) (Q_2^t)^2 }{ \alpha_2 \eta^t   \epsilon^{2}}, \label{eqn:noise_scale_WOR_A}
\end{align}\end{small}where $(a)$ holds from $\eta^t = \alpha_2  L_{W}^t $ and
\begin{small}
\begin{align}
\E \big[\|  \xib^t  \|_{F}^{2} \big]  &\overset{(a)}{=}  \frac{mk}{K} \sum_{i=1}^{K}\frac{32G^2 (Q_2^t)^2 q_{i,t}^2  \ln(1.25q_{i,t}/\delta)}{ (\eta^t)^{2} \epsilon^{2}} \notag\\
&\overset{(b)}{\leq}   \frac{32 mk G^2 (Q_2^t)^2  \ln(1.25 /\delta)}{ (\eta^t)^{2} \epsilon^{2}}. \label{eqn:total_noise_11}
\end{align}
\end{small}
In \eqref{eqn:total_noise_11},    $(a)$ follows from \eqref{eqn:noise for with replacement_0}. $(b)$ holds because of $q_{i,t}\leq 1$.
By \eqref{thm1: lip_W_1}, \eqref{thm1: lip_W_Qt} and \eqref{eqn:noise_scale_WOR_A}, for $ r \in [Q^t]\setminus [Q_1] $,  we have,
\begin{small}
\begin{align}
& \E \big[F(\ol\Wb^{t, r},\Hb^{t, r}) \big] \leq \E \big[F(\ol\Wb^{t, r-1}, \Hb^{t, r-1}) \big] +  \frac{L_{W}^t}{2} \underbrace{\E \big[ \|\ol\Wb^{t, r} - \ol\Wb^{t, r-1}\|_F^2 \big]}_{\triangleq \rm (S.2)} \notag \\
&~~~+ \underbrace{ \E \big[\langle \nabla_{W} F(\ol\Wb^{t, r-1}, \Hb^{t, r-1}), \ol\Wb^{t, r} - \ol\Wb^{t, r-1} \rangle \big]}_{\triangleq \rm (S.3)} + \frac{16 mk G^2 (Q_2^t)^2 \ln(1.25 /\delta)  }{ \alpha_2 \eta^t   \epsilon^{2}}.\label{thm1:lip_W}
\end{align}
\end{small}
The terms ${\rm (S.2)}$ and ${\rm (S.3)}$  can be bounded by  the following Lemma \ref{Lmemma:Appendix_B2} (proved in Appendix~\ref{subsec:proof of fedmavg_Lemma_1}) and  Lemma~\ref{Lmemma:Appendix_B3} (proved in Appendix~\ref{subsec:proof of fedmavg_Lemma_2}), respectively.
\begin{Lemma}\label{Lmemma:Appendix_B2}
For any $t$ and $r \in [Q^t-1]\setminus [Q_1]  $, we have
\begin{small}
\begin{align}
&\E \big[ \|\ol\Wb^{t, r} - \ol\Wb^{t, r-1}\|_F^2 \big] \leq  \frac{\phi^2}{K b (\eta^t)^2} + \frac{1}{(\eta^t)^2} \E \big[ \big\|  \frac{1}{K} \sum_{i \in \Sc^t}   \nabla_{W}F_i(\Wb_i^{t, r-1}, \Hb_i^{t, r-1})   \big\|^2 \big]. \label{thm1: descent_immediate_0}
\end{align}
\end{small}
\end{Lemma}
\begin{Lemma}\label{Lmemma:Appendix_B3}
For any $t$ and $r \in [Q^t-1]\setminus [Q_1]  $,  we have
\vspace{-0.02cm}
\begin{small}
\begin{align}
& \E \big[\langle \nabla_{W} F(\ol\Wb^{t, r-1}, \Hb^{t, r-1}), \ol\Wb^{t, r} - \ol\Wb^{t, r-1}\rangle \big]\notag \\
=& -\frac{1}{2\eta^t}  \E \Big[\big\| \nabla_{W} F(\ol\Wb^{t, r-1}, \Hb^{t, r-1}) \big\|_F^2    +  \big\| \frac{1}{K} \sum_{i \in \Sc^t}\nabla_{W}F_i(\Wb_i^{t, r-1}, \Hb_i^{t, r-1}) \big\|_F^2 \Big]\notag\\
& +  \frac{ \zeta^2}{K \eta^t}     + \frac{1}{K \eta^t} \sum_{i = 1}^{N} (L_{W_i}^t)^2\E \big[\| \ol\Wb^{t, r-1} -  \Wb_i^{t, r-1}\|_F^2 \big]. \label{thm1: descent_immediate_1}
\end{align}
\end{small}
\end{Lemma}
Thus, substituting \eqref{thm1: descent_immediate_0} and  \eqref{thm1: descent_immediate_1} into \eqref{thm1:lip_W} gives rise to
\begin{small}
\begin{align}
&\E \big[F(\ol\Wb^{t, r}, \Hb^{t, r}) \big] -\E \big[F(\ol\Wb^{t, r-1}, \Hb^{t, r-1}) \big]\notag \\
\leq&   -\frac{1}{2\eta^t}  \E \big[\big\| \nabla_{W} F(\ol\Wb^{t, r-1}, \Hb^{t, r-1}) \big\|_F^2 \big] +     \frac{L_{W}^t \phi^2}{2K b (\eta^t)^2} +    \frac{ \zeta^2}{K \eta^t} \notag\\
&+ (\frac{L_{W}^t}{2(\eta^t)^2}-  \frac{1}{2\eta^t}) \E \big[ \big\|  \frac{1}{K} \sum_{i \in \Sc^t}   \nabla_{W}F_i(\Wb_i^{t, r-1}, \Hb_i^{t, r-1})   \big\|^2 \big]  \notag \\
&  + \frac{1}{K \eta^t} \sum_{i = 1}^{N} (L_{W_i}^t)^2\E\big[\| \ol\Wb^{t, r-1} -  \Wb_i^{t, r-1}\|_F^2 \big] + \frac{16 mk G^2 (Q_2^t)^2 \ln(1.25 /\delta)  }{ \alpha_2 \eta^t   \epsilon^{2}} \notag\\
\overset{(a)}{\leq} &   -\frac{1}{2\eta^t}  \E \big[\big\| \nabla_{W} F(\ol\Wb^{t, r-1}, \Hb^{t, r-1}) \big\|_F^2 \big] +     \frac{\ol L_{W}  \phi^2}{2K b (\eta^t)^2} +    \frac{ \zeta^2}{K \eta^t} \notag\\
&  + \frac{1}{K \eta^t} \sum_{i = 1}^{N} (L_{W_i}^t)^2\E\big[ \| \ol\Wb^{t, r-1} -  \Wb_i^{t, r-1}\|_F^2 \big]  + \frac{16 mk G^2 (Q_2^t)^2 \ln(1.25 /\delta)  }{ \alpha_2 \eta^t   \epsilon^{2}},  \label{eqn: descent_Ws1}
\end{align}
\end{small}where $(a)$ follows due to $\eta^t = \alpha_2 L_{W}^t \geq L_{W}^t $ and $L_{W_i}^t \leq \ol L_W$.
Then, rearranging  the two sides of \eqref{eqn: descent_Ws1}  yields

\vspace{-0.3cm}
\begin{small}
\begin{align}
&     \E \big[\big\| \nabla_{W} F(\ol\Wb^{t, r-1}, \Hb^{t, r-1}) \big\|_F^2  \big] \notag \\
\leq & 2\eta^t \Big(\E\big[F(\ol\Wb^{t, r-1}, \Hb^{t, r-1}) \big] - \E \big[F(\ol\Wb^{t, r}, \Hb^{t, r}) \big] \Big)   \notag\\
&  + \frac{2}{K  } \sum_{i = 1}^{N} (L_{W_i}^t)^2\E \big[\| \ol\Wb^{t, r-1} -  \Wb_i^{t, r-1}\|_F^2  \big]  + \frac{\ol L_{W}  \phi^2}{ K b \eta^t} + \frac{32mk G^2  (Q_2^t)^2 \ln(1.25 /\delta) }{ \alpha_2  \epsilon^{2}}  + \frac{ 2\zeta^2}{K}. \label{eqn: descent_W_B}
\end{align}
\end{small}
\vspace{-0.3cm}

Summing \eqref{eqn: descent_W_B} up from $r =  Q_1 +1$ to $Q^t$ yields

\vspace{-0.3cm}
\begin{small}
\begin{align}
&     \sum_{r =  Q_1 +1}^{Q^t}  \E \big[ \big\| \nabla_{W} F(\ol\Wb^{t, r-1}, \Hb^{t, r-1}) \big\|_F^2  \big] \notag \\
\leq & 2\eta^t \Big(\E\big[F(\ol\Wb^{t, Q_1}, \Hb^{t, Q_1})\big] - \E\big[F(\ol\Wb^{t, Q^t}, \Hb^{t, Q^t})\big] \Big)     + \frac{2}{K  } \underbrace{\sum_{r =  Q_1 +1}^{Q^t} \sum_{i = 1}^{N} (L_{W_i}^t)^2\E \big[\| \ol\Wb^{t, r-1} -  \Wb_i^{t, r-1}\|_F^2  \big]}_{\triangleq\rm (S.4)} \notag \\
& + \frac{32mk G^2 (Q_2^t)^3 \ln(1.25 /\delta)  }{ \alpha_2  \epsilon^{2}}  +    \frac{ 2 \zeta^2  Q_2^t}{K} + \frac{ Q_2^t \ol L_{W}  \phi^2}{ K b \eta^t}. \label{eqn: descent_W_C}
\end{align}
\end{small}
\vspace{-0.3cm}

\noindent The term $\rm (S.4)$ can be bounded with the following lemma, which is proved in Appendix~\ref{subsec:proof of fedmavg_Lemma_3}.

\vspace{-0.3cm}
\begin{Lemma}\label{Lmemma:Appendix_3}
Let $\alpha_2 \geq   Q_2^t \sqrt{ 3 (1+ \ol L_W^2/\underline{L}_W^2 )}$. For any $t$ and $r \in [Q^t]\setminus [Q_1]$, it holds that

\vspace{-0.3cm}
{\small \begin{align}
\sum_{r =  Q_1 +1}^{Q^t}\sum_{i = 1}^{N}(L_{W_i}^t)^2\E[\| \ol\Wb^{t, r-1} -  \Wb_i^{t, r-1}\|_F^2] \leq  \frac{4 N \zeta^2 C_1^t  }{K \alpha_2^2}, \label{eqn: descent_W_DD}
\end{align}}where $C_1^t \triangleq    Q_2^t(Q_2^t - 1)(2Q_2^t - 1)$.
\end{Lemma}	
By applying Lemma \ref{Lmemma:Appendix_3} and plugging \eqref{eqn: descent_W_DD} into \eqref{eqn: descent_W_C}, we have

\vspace{-0.5cm}
\begin{small}
\begin{align}
& \sum_{r =  Q_1 +1}^{Q^t} \E \big[ \big\| \nabla_{W} F(\ol\Wb^{t, r-1}, \Hb^{t, r-1}) \big\|_F^2  \big] \notag \\
\leq & 2\eta^t  \Big(\E\big[F(\ol\Wb^{t, Q_1}, \Hb^{t, Q_1})] - \E[F(\ol\Wb^{t, Q^t}, \Hb^{t, Q^t})\big] \Big)   \notag\\
& + \frac{32mk G^2    (Q_2^t)^3 \ln(1.25 /\delta)  }{ \alpha_2  \epsilon^{2}} +    \frac{ 2 \zeta^2  Q_2^t}{K}   + \frac{ Q_2^t \ol L_{W}  \phi^2}{ K b \eta^t} +   \frac{8 N \zeta^2 C_1^t  }{ K^2 \alpha_2^2}. \label{eqn: descent_W_D}
\end{align}
\end{small}
\vspace{-0.3cm}

Combining \eqref{thm1: descent_H2} and \eqref{eqn: descent_W_D}  yields

\vspace{-0.4cm}
{\small \begin{align}
& \sum_{r = 1}^{Q_1} \sum_{i = 1}^{N} \E\big[\|\Hb_i^{t, r-1} - \Hb_i^{t, r}\|_F^2\big] + \sum_{r =  Q_1 +1}^{Q^t} \E \big[ \big\| \nabla_{W} F(\ol\Wb^{t, r-1}, \Hb^{t, r-1}) \big\|_F^2  \big] \notag \\
\overset{(a)}{\leq} & 2\eta^t \Big(\E\big[F(\ol\Wb^{t, 0}, \Hb^{t, 0})] -\E[F(\ol\Wb^{t, Q^t}, \Hb^{t, Q^t})\big] \Big)\notag\\
& + \frac{32mk G^2  (Q_2^t)^3 \ln(1.25 /\delta)  }{ \alpha_2  \epsilon^{2}} +    \frac{ 2 \zeta^2  Q_2^t}{K}   + \frac{ Q_2^t \ol L_{W}  \phi^2}{ K b \eta^t}  +   \frac{8 N \zeta^2 C_1^t  }{ K^2 \alpha_2^2}, \label{thm1: descent_WH}
\end{align}}where $(a)$ holds because of $\eta^t \geq 1/(( \alpha_1 - 1) \ol L_H)$.

\vspace{0.2cm}
\noindent ({\bf Derivation of the Main Result})
We next derive the convergence in terms of the optimal gap functions in \eqref{eqn: prox_H} and \eqref{eqn: prox_W}.
From \eqref{thm1: descent_WH}  and   $\gamma_i^{t} =  \alpha_1 L_H^t/2  $ and $\eta^t = \alpha_2 L_{W}^t$, we have

\vspace{-0.4cm}
{\small
\begin{align}
&\!\!\!\!\!\! \sum_{r = 1}^{Q_1}\E[G_{H}(\ol \Wb^{t,r-1}, \Hb^{t,r-1})] = \sum_{r = 1}^{Q_1}\sum_{i=1}^{N} (\gamma_i^{t})^2 \E[\|\Hb_i^{t, r-1} - \Hb_i^{t, r}\|_F^2] \notag \\
\overset{(a)}{\leq} & 2 \alpha_1^2 \ol L_H^2 \Big(  \alpha_2  \ol L_W \E[F(\ol\Wb^{t, 0}, \Hb^{t, 0})] -\E[F(\ol\Wb^{t, Q^t}, \Hb^{t, Q^t})] \notag\\
& + \frac{16  mk G^2  (Q_2^t)^3 \ln(1.25 /\delta)  }{ \alpha_2  \epsilon^{2}} +    \frac{ \zeta^2  Q_2^t}{K}  + \frac{ Q_2^t \ol L_{W}  \phi^2}{2 K b \underline{L}_W \alpha_2 }  +   \frac{4 N \zeta^2 C_1^t  }{ K^2 \alpha_2^2} \Big), \label{eqn: descent_WH}
\end{align}}where $(a)$ follows because $\gamma_i^{t} \leq  \alpha_1 \ol L_H/2  $ and $  \alpha_2 \underline{L}_W \leq  \eta^t \leq  \alpha_2 \ol L_W$.
Then, summing   \eqref{eqn: descent_WH}  up from $t = 1$ to $R$ yields
{\small
\begin{align}
& \sum_{t = 1}^{R}\sum_{r = 1}^{Q_1} \E\big[G_{H}(\ol \Wb^{t,r-1}, \Hb^{t,r-1})\big] \notag \\
\leq &  2 \alpha_1^2 \ol L_H^2 \Big(  \alpha_2  \ol L_W \big(F(\ol\Wb^{1, 0}, \Hb^{1, 0}) - \underline{F}\big)  + \frac{16  mk G^2  \ln(1.25 /\delta)   \sum_{t = 1}^{R} (Q_2^t)^3 }{ \alpha_2  \epsilon^{2}} +    \frac{ \zeta^2 \sum_{t = 1}^{R} Q_2^t}{K}    \notag\\
& + \frac{  \ol L_{W}  \phi^2 \sum_{t = 1}^{R} Q_2^t }{2 K b \underline{L}_W \alpha_2 }  +   \frac{4 N \zeta^2 \sum_{t = 1}^{R} C_1^t}{ K^2 \alpha_2^2} \Big). \label{thm1: descent_Hs_irox}
\end{align}}
Similarly, from \eqref{thm1: descent_WH}, we have
\vspace{-0.2cm}
{\small \begin{align}
&\sum_{t = 1}^{R}\sum_{r =  Q_1 + 1}^{Q^t} \E \big[ \big\| \nabla_{W} F(\ol\Wb^{t, r-1}, \Hb^{t, r-1}) \big\|_F^2  \big]  \notag \\
=&\sum_{t = 1}^{R}\sum_{r =  Q_1 + 1}^{Q^t}\E \big[G_{W}(\ol \Wb^{t,r-1}, \Hb^{t, r - 1})] \big]  \notag \\
\leq &   2 \alpha_2  \ol L_W \big(F(\ol\Wb^{1, 0}, \Hb^{1, 0}) - \underline{F}\big)  + \frac{32mk G^2  \ln(1.25 /\delta)   \sum_{t = 1}^{R} (Q_2^t)^3 }{ \alpha_2  \epsilon^{2}} +    \frac{ 2  \zeta^2 \sum_{t = 1}^{R} Q_2^t }{K}  \notag\\
& + \frac{  \ol L_{W}  \phi^2 \sum_{t = 1}^{R} Q_2^t }{ K b \underline{L}_W \alpha_2 }  +   \frac{8 N   \zeta^2  \sum_{t = 1}^{R} C_1^t}{ K^2 \alpha_2^2}. \label{thm1: descent_Ws2}
\end{align}}\vspace{-0.3cm}

\noindent  By combining \eqref{thm1: descent_Hs_irox} and \eqref{thm1: descent_Ws2}, and then dividing two sides of summation result  by $T=RQ_1+ \sum_{t = 1}^{R} Q_2^t$ yields

\vspace{-0.3cm}
\begin{small}
\begin{align}
&\!\!\! \frac{1}{T}\Big[\sum_{t = 1}^{R}\sum_{r = 1}^{Q_1} \E\big[G_{H}(\ol \Wb^{t,r-1}, \Hb^{t, r-1})\big] +\sum_{t = 1}^{R}\sum_{r =  Q_1 + 1}^{Q^t}\E\big[G_{W}(\ol \Wb^{t,r-1}, \Hb^{t, r - 1})\big] \Big]\notag \\
\leq &  \frac{2 (\alpha_1^2 \ol L_H^2 + 1)}{T} \Big(  \alpha_2  \ol L_W \big(F(\ol\Wb^{1, 0}, \Hb^{1, 0}) - \underline{F}\big)  + \frac{16  mk G^2  \ln(1.25 /\delta)   \sum_{t = 1}^{R} (Q_2^t)^3 }{ \alpha_2  \epsilon^{2}}  +    \frac{ \zeta^2 \sum_{t = 1}^{R} Q_2^t}{K}    \notag\\
& + \frac{\ol L_{W}  \phi^2 \sum_{t = 1}^{R} Q_2^t }{2 K b \underline{L}_W \alpha_2 }  +   \frac{4 N \zeta^2 \sum_{t = 1}^{R} C_1^t}{ K^2 \alpha_2^2} \Big).
\end{align}
\end{small}
\noindent This completes the proof.  \hfill $\blacksquare$

\section{Proofs of Key Lemmas for Theorem \ref{thm: model_avg} }\label{sec:proof of fedmavg_Lemma}
\subsection{Proof of Lemma \ref{Lmemma:Appendix_B2}} \label{subsec:proof of fedmavg_Lemma_1}
According to \eqref{eqn: q_1_appendix}, we have
\begin{small}
\begin{align}
\!\!\!\!&\E \big[ \|\ol\Wb^{t, r} - \ol\Wb^{t, r-1}\|_F^2 \big] \notag\\
=& \frac{1}{(\eta^t)^2} \E \big[ \big\|\frac{1}{K} \sum_{i \in \Sc^t} \nabla_{W}F_i(\Wb_i^{t, r-1}, \Hb_i^{t, r-1};\mathcal{B}_i^{t,r}) \big\|^2\big] \notag\\
\overset{(a)}{=}& \frac{1}{(\eta^t)^2} \E \Big[ \big\|  \frac{1}{K} \sum_{i \in \Sc^t} \big(\nabla_{W}F_i(\Wb_i^{t, r-1}, \Hb_i^{t, r-1};\mathcal{B}_i^{t,r}) - \nabla_{W}F_i(\Wb_i^{t, r-1}, \Hb_i^{t, r-1}) \big) \big\|^2 \Big] \notag\\
&+ \frac{1}{(\eta^t)^2} \E \big[ \big\|  \frac{1}{K} \sum_{i \in \Sc^t}   \nabla_{W}F_i(\Wb_i^{t, r-1}, \Hb_i^{t, r-1})   \big\|^2 \big] \notag\\
\overset{(b)}{=}& \frac{1}{(\eta^t)^2 K^2} \E \Big[ \sum_{i \in \Sc^t}  \big\|  \nabla_{W}F_i(\Wb_i^{t, r-1}, \Hb_i^{t, r-1};\mathcal{B}_i^{t,r}) - \nabla_{W}F_i(\Wb_i^{t, r-1}, \Hb_i^{t, r-1})  \big\|^2 \Big] \notag\\
&+ \frac{1}{(\eta^t)^2} \E \big[ \big\|  \frac{1}{K} \sum_{i \in \Sc^t}   \nabla_{W}F_i(\Wb_i^{t, r-1}, \Hb_i^{t, r-1})   \big\|^2 \big] \notag\\
\overset{(c)}{\leq}& \frac{1}{(\eta^t)^2} \E \big[ \big\|  \frac{1}{K} \sum_{i \in \Sc^t}   \nabla_{W}F_i(\Wb_i^{t, r-1}, \Hb_i^{t, r-1})   \big\|^2 \big]
+   \frac{\phi^2}{K b (\eta^t)^2},
\end{align}
\end{small}where $(a)$ follows because $\E[\|\mathbf{Z}\|^2]=\E[\|\mathbf{Z}-\E[\mathbf{Z}]\|^2]+\|\E[\mathbf{Z}]\|^2$; $(b)$ follows because   $\nabla_{W}F_i(\Wb_i^{t, r-1}, \Hb_i^{t, r-1};\mathcal{B}_i^{t,r})  - \nabla_{W}F_i(\Wb_i^{t, r-1}, \Hb_i^{t, r-1})$ is independent across  the clients;  $(c)$ holds due to Assumption \ref{Ass: bounded_Gradient_variance}.  \hfill $\blacksquare$

\subsection{Proof of Lemma \ref{Lmemma:Appendix_B3}} \label{subsec:proof of fedmavg_Lemma_2}

\vspace{-0.4cm}
\begin{small}
\begin{align}
& \E \big[\langle \nabla_{W} F(\ol\Wb^{t, r-1}, \Hb^{t, r-1}), \ol\Wb^{t, r} - \ol\Wb^{t, r-1}\rangle \big]\notag \\
\overset{(a)}{=} & -\frac{1}{\eta^t} \E \Big[\big\langle \nabla_{W} F(\ol\Wb^{t, r-1}, \Hb^{t, r-1}),  \frac{1}{K} \sum_{i \in \Sc^t}\nabla_{W}F_i(\Wb_i^{t, r-1}, \Hb_i^{t, r-1};\mathcal{B}_i^{t,r}) \big\rangle \Big]\notag\\
\overset{(b)}{=} & -\frac{1}{\eta^t} \E \Big[\big\langle \nabla_{W} F(\ol\Wb^{t, r-1}, \Hb^{t, r-1}),  \frac{1}{K} \sum_{i \in \Sc^t}\nabla_{W}F_i(\Wb_i^{t, r-1}, \Hb_i^{t, r-1}) \big\rangle \Big]\notag\\
\overset{(c)}{=}& -\frac{1}{2\eta^t}  \E \Big[\big\| \nabla_{W} F(\ol\Wb^{t, r-1}, \Hb^{t, r-1}) \big\|_F^2 \Big]  -\frac{1}{2\eta^t}  \E \Big[ \big\| \frac{1}{K} \sum_{i \in \Sc^t}\nabla_{W}F_i(\Wb_i^{t, r-1}, \Hb_i^{t, r-1}) \big\|_F^2 \Big] \notag \\
&+ \frac{1}{2\eta^t}  \E \Big[ \big\| \nabla_{W} F(\ol\Wb^{t, r-1}, \Hb^{t, r-1}) - \frac{1}{K} \sum_{i \in \Sc^t}\nabla_{W}F_i(\Wb_i^{t, r-1}, \Hb_i^{t, r-1})  \big\|_F^2 \Big],    \label{eqn: bound_S1}
\end{align}
\end{small}where $(a)$ holds due to \eqref{eqn: FedAM update of W2 0}; $(b)$ follows from Assumption~\ref{Ass: bounded_Gradient_variance};   $(c)$ follows from the basic identity $\langle\mathbf{Z}_1, \mathbf{Z}_2 \rangle=\frac{1}{2}(\|\mathbf{Z}_1\|^2+\|\mathbf{Z}_2\|^2-\|\mathbf{Z}_1-\mathbf{Z}_2\|^2)$.

The last term  in \eqref{eqn: bound_S1}  can be further bounded by
\begin{small}
\begin{align}
&\E \Big[\big\| \nabla_{W} F(\ol\Wb^{t, r-1}, \Hb^{t, r-1})  - \frac{1}{K} \sum_{i \in \Sc^t}\nabla_{W}F_i(\Wb_i^{t, r-1}, \Hb_i^{t, r-1})  \big\|_F^2 \Big]\notag\\
=& \E \Big[ \big\| \frac{1}{K} \sum_{i \in \Sc^t} \big(\nabla_{W} F(\ol\Wb^{t, r-1}, \Hb^{t, r-1}) -   \nabla_{W} F_i(\ol\Wb^{t, r-1}, \Hb^{t, r-1}) \notag\\
&+ \nabla_{W} F_i(\ol\Wb^{t, r-1}, \Hb^{t, r-1}) - \nabla_{W}F_i(\Wb_i^{t, r-1}, \Hb_i^{t, r-1})  \big)\big\|_F^2 \Big]\notag\\
\leq& \frac{2}{K^2} \E \Big[ \big\|  \sum_{i \in \Sc^t} \big(\nabla_{W} F(\ol\Wb^{t, r-1}, \Hb^{t, r-1}) - \nabla_{W} F_i(\ol\Wb^{t, r-1}, \Hb^{t, r-1})  \big)\big\|_F^2 \Big] \notag\\
&+ \frac{2}{K^2} \E \Big[ \big\|  \sum_{i \in \Sc^t} \big( \nabla_{W} F_i(\ol\Wb^{t, r-1}, \Hb^{t, r-1}) - \nabla_{W}F_i(\Wb_i^{t, r-1}, \Hb_i^{t, r-1})  \big)\big\|_F^2 \Big] \notag\\
\overset{(a)}{\leq} &  \frac{2\zeta^2}{K}     + \frac{2}{K}   \E \big[ \sum_{i \in \Sc^t} (L_{W_i}^t)^2 \| \ol\Wb^{t, r-1} -  \Wb_i^{t, r-1}\|_F^2 \big] \notag\\
\leq &  \frac{2\zeta^2}{K}     + \frac{2}{K}   \sum_{i = 1}^{N} (L_{W_i}^t)^2\E\big[\| \ol\Wb^{t, r-1} -  \Wb_i^{t, r-1}\|_F^2\big],  \label{eqn: bound_c}
\end{align}
\end{small}where the first term   in the RHS of $(a)$ comes from  Assumption~\ref{Ass: non_iid}, and the   second term   in the RHS of \eqref{lemS1: bound_c} follows because of Assumption~\ref{Ass: bounded_Lipschitz}.
Then,  Plugging \eqref{eqn: bound_c}  into \eqref{eqn: bound_S1} yields
\begin{small}
\begin{align}
& \E \big[\langle \nabla_{W} F(\ol\Wb^{t, r-1}, \Hb^{t, r-1}), \ol\Wb^{t, r} - \ol\Wb^{t, r-1}\rangle \big]\notag \\
\leq& -\frac{1}{2\eta^t}  \E \big[\big\| \nabla_{W} F(\ol\Wb^{t, r-1}, \Hb^{t, r-1}) \big\|_F^2  \big]-\frac{1}{2\eta^t}  \E \big[ \big\| \frac{1}{K} \sum_{i \in \Sc^t}\nabla_{W}F_i(\Wb_i^{t, r-1}, \Hb_i^{t, r-1}) \big\|_F^2  \big] \notag \\
&+  \frac{ \zeta^2}{K \eta^t}     + \frac{1}{K \eta^t} \sum_{i = 1}^{N} (L_{W_i}^t)^2\E\big[\| \ol\Wb^{t, r-1} -  \Wb_i^{t, r-1}\|_F^2\big].
\end{align}
\end{small}
Thus, we complete the proof. \hfill $\blacksquare$

\subsection{Proof of Lemma \ref{Lmemma:Appendix_3}} \label{subsec:proof of fedmavg_Lemma_3}
According to the definition of $\ol\Wb^{t, r-1}$, for $\forall r \in [Q^t]\setminus [Q_1]$, we have
\begin{small}
\begin{align}
\ol\Wb^{t, r-1} &= \frac{1}{K} \sum\limits_{i \in \Sc^t}  \Wb_i^{t, r-1}\notag  \\
& \overset{(a)}{=} \frac{1}{K}  \sum_{i \in \Sc^t} \Big(\Wb^{t} -\frac{1}{\eta^t} \sum_{j = Q_1}^{r-1} \nabla_{W}F_i(\Wb_i^{t, j-1}, \Hb_i^{t, j-1}; \mathcal{B}_i^{t,j})\Big) \notag \\
& = \Wb^{t} - \frac{1}{\eta^t K}\sum_{j = Q_1}^{r-1} \sum_{i \in \Sc^t}  \nabla_{W}F_i(\Wb_i^{t, j-1},\Hb_i^{t, j-1}; \mathcal{B}_i^{t,j}), \label{lem2: W_def}
\end{align}
\end{small}
where $(a)$ is obtained by applying \eqref{eqn: FedAM update of W2 0}, that is
\begin{small}
\begin{align}\label{lem2: W_def5}
\Wb_i^{t,r-1} = \Wb^{t} - \frac{1}{\eta^t}\sum_{j = Q_1}^{r-1} \nabla_{W} F_i(\Wb_i^{t, j-1}, \Hb_i^{t, j-1}; \mathcal{B}_i^{t,j}).
\end{align}
\end{small}
As a result, by \eqref{lem2: W_def} and \eqref{lem2: W_def5}, we have
\begin{small}
\begin{align}
&\E\big[\|\ol\Wb^{t, r-1} - \Wb_i^{t, r-1}\|_F^2\big] \notag \\
= &\E\Big[\Big\|\Wb^{t} - \frac{1}{\eta^t K}\sum_{j = Q_1}^{r-1} \sum_{i \in \Sc^t} \nabla_{W}F_i(\Wb_i^{t, j-1}, \Hb_i^{t, j-1}; \mathcal{B}_i^{t,j}) - \big(\Wb^{t} - \frac{1}{\eta^t}\sum_{j = Q_1}^{r-1} \nabla_{W} F_i(\Wb_i^{t, j-1}, \Hb_i^{t, j-1}; \mathcal{B}_i^{t,j})\big)\Big\|_F^2 \Big] \notag  \\
= &  \frac{1}{(\eta^t)^2}\E\Big[\Big\|\frac{1}{K} \sum_{j = Q_1}^{r-1} \sum_{i \in \Sc^t} \nabla_{W}F_i(\Wb_i^{t, j-1}, \Hb_i^{t, j-1}; \mathcal{B}_i^{t,j}) - \sum_{j = Q_1}^{r-1} \nabla_{W} F_i(\Wb_i^{t, j-1}, \Hb_i^{t, j-1}; \mathcal{B}_i^{t,j})\Big\|_F^2\Big] \notag  \\
\leq &\frac{(r -Q_1)}{(\eta^t)^2}\sum_{j = Q_1}^{r-1}\E\Big[\Big\| \frac{1}{K} \sum_{i \in \Sc^t} \nabla_{W}F_i(\Wb_i^{t, j-1}, \Hb_i^{t, j-1}; \mathcal{B}_i^{t,j}) - \nabla_{W} F_k(\Wb_k^{t, j-1}, \Hb_k^{t, j-1}; \mathcal{B}_k^{t,j})\Big\|_F^2 \Big] \notag \\
\overset{(b)}{=} &\frac{(r -Q_1)}{(\eta^t)^2}\sum_{j = Q_1}^{r-1} \Big\| \frac{1}{K} \sum_{i \in \Sc^t} \Big(\nabla_{W}F_i(\Wb_i^{t, j-1}, \Hb_i^{t, j-1} ) - \nabla_{W} F_k(\Wb_k^{t, j-1}, \Hb_k^{t, j-1} ) \Big)\Big\|_F^2  \notag \\
\overset{(c)}{\leq} &\frac{(r -Q_1)}{(\eta^t)^2 K}\sum_{j = Q_1}^{r-1}  \sum_{i=1}^{N}  \Big\|\nabla_{W}F_i(\Wb_i^{t, j-1}, \Hb_i^{t, j-1} ) - \nabla_{W} F_k(\Wb_k^{t, j-1}, \Hb_k^{t, j-1} )\Big\|_F^2, \label{lem2: bound_diff1}
\end{align}
\end{small}where $(b)$   holds since $\nabla_{W}F_i(\Wb_i^{t, j-1}, \Hb_i^{t, j-1}; \mathcal{B}_i^{t,j})  - \nabla_{W} F_k(\Wb_k^{t, j-1}, \Hb_k^{t, j-1}; \mathcal{B}_k^{t,j})$ is independent across the clients;  $(c)$  follows by using the inequality $\|\sum_{i=1}^N \mathbf{z}_i\|^2 \leq   N \sum_{i=1}^N\left\|\mathbf{z}_i\right\|^2$ for any vectors $\mathbf{z}_i$ and any positive integer $N$.
Then, the term $\|\nabla_{W}F_i(\Wb_i^{t, j-1}, \Hb_i^{t, j-1} )-\nabla_{W} F_k(\Wb_k^{t, j-1}, \Hb_k^{t, j-1} )\|_F^2$ in \eqref{lem2: bound_diff1}  can be   bounded by
\begin{small}
\begin{align}
& \big\|\nabla_{W}F_i(\Wb_i^{t, j-1}, \Hb_i^{t, j-1}) - \nabla_{W} F_k(\Wb_k^{t, j-1}, \Hb_k^{t, j-1})\big\|_F^2 \notag \\
\leq & \Big\|\nabla_{W}F_i(\Wb_i^{t, j-1}, \Hb_i^{t, j-1}) - \nabla_{W}F_i(\ol \Wb^{t, j-1}, \Hb_i^{t, j-1}) \notag \\
&+ \nabla_{W}F_i(\ol \Wb^{t, j-1}, \Hb_i^{t, j-1}) - \nabla_{W}F(\ol \Wb^{t, j-1}, \Hb^{t, j-1}) \notag\\
&+  \nabla_{W}F(\ol \Wb^{t, j-1}, \Hb^{t, j-1})- \Big(\nabla_{W}F_k(\Wb_k^{t,j-1}, \Hb_k^{t, j-1})\notag \\
& - \nabla_{W}F_k(\ol \Wb^{t,j-1}, \Hb_k^{t, j-1}) + \nabla_{W}F_k(\ol \Wb^{t, j-1}, \Hb_k^{t, j-1})\notag \\
& - \nabla_{W}F(\ol \Wb^{t,j-1}, \Hb^{t, j-1}) + \nabla_{W}F(\ol \Wb^{t, j-1}, \Hb^{t, j-1})\Big)\Big\|_F^2 \notag \\
\leq& 4\|\nabla_{W}F_i(\Wb_i^{t, j-1}, \Hb_i^{t, j-1}) - \nabla_{W}F_i(\ol \Wb^{t, j-1}, \Hb_i^{t, j-1})\|_F^2 \notag \\
&+ 4\|\nabla_{W}F_i(\ol \Wb^{t, j-1}, \Hb_i^{t, j-1}) - \nabla_{W}F(\ol \Wb^{t, j-1}, \Hb^{t, j-1})\|_F^2 \notag \\
&+ 4\|\nabla_{W}F_k(\Wb_k^{t, j-1}, \Hb_k^{t, j-1}) - \nabla_{W}F_k(\ol \Wb^{t, j-1}, \Hb_k^{t, j-1})\|_F^2 \notag \\
&+ 4\|\nabla_{W}F_k(\ol \Wb^{t, j-1}, \Hb_k^{t, j-1}) - \nabla_{W}F(\ol \Wb^{t, j-1}, \Hb^{t, j-1})\|_F^2 \notag \\
\overset{(d)}{\leq} &4(L_{W_i}^t)^2\|\ol \Wb^{t, j-1} - \Wb_i^{t, j-1}\|_F^2  + 4(L_{W_k}^t)^2\|\ol \Wb^{t, j-1} - \Wb_k^{t, j-1}\|_F^2 + 8\zeta^2, \label{lemS1: bound_c}
\end{align}
\end{small}where $(d)$ follows from Assumption~\ref{Ass: non_iid}.
Then, substituting \eqref{lemS1: bound_c} into \eqref{lem2: bound_diff1} gives rise to
\begin{small}
\begin{align}
&\sum_{r =  Q_1+1}^{Q^t}\sum_{i = 1}^{N}   (L_{W_i}^{t})^2  \E\big[\|\ol\Wb^{t, r-1} - \Wb_i^{t, r-1}\|_F^2\big] \notag\\
\leq&~\sum_{r =  Q_1+1}^{Q^t}\sum_{i = 1}^{N}   (L_{W_i}^{t})^2  \Big(\frac{ (r -Q_1-1)}{K(\eta^t)^2}\sum_{j = Q_1}^{r-2}\sum_{i = 1}^{N}  \Big(4(L_{W_i}^t)^2\|\ol\Wb^{t, j-1}- \Wb_i^{t, j-1}\|_F^2 \notag\\
&  + 8\zeta^2   + 4(L_{W_k}^t)^2\|\ol\Wb^{t, j-1} - \Wb_k^{t, j-1}\|_F^2\Big) \Big) \notag \\
\overset{(e)}{=}& \frac{N}{K}\sum_{r =  Q_1+1}^{Q^t}  \frac{4(r -Q_1-1)}{ (\eta^t/L_W^t)^2}\sum_{j = Q_1}^{r-2}\sum_{i = 1}^{N}  (L_{W_i}^t)^2 \|\ol\Wb^{t, j-1} - \Wb_i^{t, j-1}\|_F^2 + \frac{N}{K}\sum_{r =  Q_1+1}^{Q^t}  \frac{8(r -Q_1-1)^2}{ (\eta^t/L_W^t)^2}\zeta^2   \notag \\
&+ \frac{N}{K}\sum_{r =  Q_1+1}^{Q^t}   \frac{4(r - Q_1-1)}{(\eta^t/L_W^t)^2} \sum_{j=Q_1}^{r-2} \sum_{i = 1}^{N} (L_{W_i}^{t})^2 \cdot \big(\frac{L_{W_i}^t}{L_{W}^t}\big)^2\|\ol\Wb^{t,j-1} - \Wb_i^{t, j-1}\|_F^2 \notag\\
\overset{(f)}{\leq}& \frac{2NQ_2^t(Q_2^t - 1)}{K\alpha_2^2}(1+ \frac{\ol L_W^2}{\underline{L}_W^2})\sum_{r =  Q_1 +1}^{Q^t}\sum_{i=1}^{N} (L_{W_i}^t)^2 \big\|\ol\Wb^{t, j-1} - \Wb_i^{t, j-1} \big\|_F^2  +   \frac{4 N Q_2^t(Q_2^t - 1)(2Q_2^t - 1)  \zeta^2  }{ 3 K\alpha_2^2}, \label{eqn: c_bound1}
\end{align}
\end{small}where $(e)$ follows since $(L_W^t)^2 = (1/N)\sum_{i = 1}^{N}(L_{W_i}^t)^2$; $(f)$ follows due to   $\frac{(L_{W_i}^t)^2}{(L_W^t)^2} \leq \frac{\ol L_W^2}{\underline{L}_W^2}$ and $\eta^t = \alpha_2 L_W^t$, and
\begin{small}
\begin{align}
\sum_{r=Q_1 + 1}^{Q^t} (r-1-Q_1)\sum_{j = Q_1}^{r -2}a_j
\leq \sum_{r=Q_1 + 1}^{Q^t} \frac{Q_2^t(Q_2^t-1)}{2} a_{r-1}, \forall a_j > 0, \label{thm1: e_bound4}
\end{align}
\end{small}
and
\begin{small}
\begin{align}
\sum_{r =  Q_1 + 1}^{Q^t} (r-1 - Q_1 )^2 = \frac{Q_2^t(Q_2^t - 1)(2Q_2^t - 1)}{6} \label{eqn: square_sum}.
\end{align}
\end{small}
Since $\alpha_2 \geq   Q_2^t \sqrt{ 3 (1+ \ol L_W^2/\underline{L}_W^2 )}$, implies  $\alpha_2^2 \geq 2Q_2^t(Q_2^t - 1)(1+ \ol L_W^2/\underline{L}_W^2 )$. After rearranging \eqref{eqn: c_bound1}, we obtain
\begin{small}
\begin{align}
\sum_{r =  Q_1+1}^{Q^t}\sum_{i = 1}^{N}  (L_{W_i}^{t})^2 \E\big[\|\ol\Wb^{t, r-1}- \Wb_i^{t, r-1}\|_F^2\big] \leq&~ \frac{4 N Q_2^t(Q_2^t - 1)(2Q_2^t - 1)  \zeta^2}{3K\big(\alpha_2^2 - 2Q_2^t(Q_2^t - 1)(1+ \ol L_W^2/\underline{L}_W^2 ) } \notag\\
\overset{(g)}{\leq}&~ \frac{4 N Q_2^t(Q_2^t - 1)(2Q_2^t - 1)  \zeta^2}{ K\alpha_2^2   } \notag\\
=& \frac{4 N C_1^t \zeta^2}{K\alpha_2^2},
\end{align}
\end{small}where $C_1^t = Q_2^t(Q_2^t - 1)(2Q_2^t - 1)$,   $(g)$ follows since $\alpha_2^2 - 2Q_2^t(Q_2^t - 1)(1+ \ol L_W^2/\underline{L}_W^2 ) > \alpha_2^2/3$.    \hfill $\blacksquare$

\section{Per-iteration complexity of Algorithm~\ref{alg: model_avg} }\label{sec:proof Lemma6}
According to \eqref{eqn: FedAM update of H1 0} and \eqref{eqn: FedAM update of W2 0}, let us revisit  $\Hb_i^{t, r}$ and $\Wb_i^{t, r}$  as follows
\begin{align}
\Hb_i^{t, r}   &=    \Big[ \Hb_i^{t,r-1}\! -\! \frac{1}{\gamma_i^{t}}{\nabla_{H_i}F_i(\Wb^{t-1}, \Hb_i^{t,r-1})} \Big]^{+}, \label{eqn: H update}\\
\Wb_i^{t, r} &= \Wb_i^{t, r-1} - \frac{1}{\eta^t}{\nabla_{W}F_i(\Wb_i^{t, r-1}, \Hb_i^{t, Q_i};\mathcal{B}_i^{t,r})}.\label{eqn: W update}
\end{align}
For simplicity, we omit the outer iteration number $t$ and inner iteration number $r$.  By the definition of $F_i(\Wb, \Hb_i)$ in \eqref{eqn: obj of client p}, $\nabla_{H_i}F_i(\Wb_i, \Hb_i)$ and $\nabla_{W}F_i(\Wb_i, \Hb_i;\mathcal{B}_i)$ can be computed as
\begin{align}
\nabla_{H_i}F_i (\Wb_i, \Hb_i) =& 2 \Wb_i^T\Wb_i \Hb_i - 2 \Wb_i^T \Xb_i  + 2\rho \oneb\oneb^T \Hb_i + (\mu_h - \rho)\Hb_i \label{eqn: gradient_H}, \\
\nabla_{W}F_i(\Wb_i, \Hb_i;\mathcal{B}_i) =& 2 \Wb_i  \Hb_i  \Hb_i^T  - 2   \Xb_i \Hb_i^T  + \mu_w \Wb_i. \label{eqn: gradient_W}
\end{align}
Thus,  the complexity order of computing $\Hb_i^{t,r}$  (mainly due to \eqref{eqn: gradient_H}) at each client $i$ can be estimated as
\begin{align}\label{eqn:HH_complexity}
(m k^2  +  k^2 n_i   + k n_i) +  (m k n_i + n_i k) + n_i k^2 +  2 n_i k  \implies  \mathcal{O}((m + n_i)k^2 + mn_ik),
\end{align}
and that  of  computing $\Wb_i^{t,r}$ (mainly due to \eqref{eqn: gradient_W}) as
\begin{align}\label{eqn:WW_complexity}
(m k b +  m k b +mk ) + (m k b + mk )  + m k \implies   \mathcal{O}(m  k b).
\end{align}
Because the complexity of $\Wb_i^{t,r}$ is much smaller than that of $\Hb_i^{t,r}$ (due to $b=|{\cal B}_i|<n_i\ll n=\sum_i^N n_i$), the total complexity order of updating $\Hb_i^{t,r}$ and $\Wb_i^{t,r}$ can be approximated by   that of updating $\Hb_i^{t,r}$. As a result, provided that all the $N$  clients (the worst case) join the learning process, one can obtain the total complexity order as $\mathcal{O}\left((mN + n)k^2 + m n k \right)$ at the client side. Moreover, the complexity order at the PS side is simply $\mathcal{O}(mkN)$.

\bibliographystyle{IEEEtran}
\bibliography{reference}

\begin{thebibliography}{10}
\providecommand{\url}[1]{#1}
\csname url@samestyle\endcsname
\providecommand{\newblock}{\relax}
\providecommand{\bibinfo}[2]{#2}
\providecommand{\BIBentrySTDinterwordspacing}{\spaceskip=0pt\relax}
\providecommand{\BIBentryALTinterwordstretchfactor}{4}
\providecommand{\BIBentryALTinterwordspacing}{\spaceskip=\fontdimen2\font plus
\BIBentryALTinterwordstretchfactor\fontdimen3\font minus
  \fontdimen4\font\relax}
\providecommand{\BIBforeignlanguage}[2]{{%
\expandafter\ifx\csname l@#1\endcsname\relax
\typeout{** WARNING: IEEEtran.bst: No hyphenation pattern has been}%
\typeout{** loaded for the language `#1'. Using the pattern for}%
\typeout{** the default language instead.}%
\else
\language=\csname l@#1\endcsname
\fi
#2}}
\providecommand{\BIBdecl}{\relax}
\BIBdecl

\bibitem{wang2021federated}
Z.~Wang, X.~Wang, R.~Sun, and T.-H. Chang, ``Federated semi-supervised learning
  with class distribution mismatch,'' \emph{arXiv preprint arXiv:2111.00010},
  2021.

\bibitem{McMahan2016FederatedLO}
H.~McMahan, E.~Moore, D.~Ramage, and B.~A. y~Arcas, ``Federated learning of
  deep networks using model averaging,'' \emph{arXiv preprint
  arXiv:1602.05629}, 2016.

\bibitem{li2022federated}
Y.~Li, S.~Wang, T.-H. Chang, and C.-Y. Chi, ``Federated stochastic primal-dual
  learning with differential privacy,'' \emph{arXiv preprint arXiv:2204.12284},
  2022.

\bibitem{CE_DDNN_2017}
H.~B. McMahan, E.~Moore, D.~Ramage, S.~Hampson, and B.~A. Areas,
  ``Communication-efficient learning of deep networks from decentralized
  data,'' in \emph{Proc. International Conference on Machine Learning (ICML)},
  2017, pp. 1--10.

\bibitem{li2022secure}
S.~Li, S.~Hou, B.~Buyukates, and S.~Avestimehr, ``Secure federated
  clustering,'' \emph{arXiv preprint arXiv:2205.15564}, 2022.

\bibitem{Parallel_RSGD_2019}
H.~Yu, S.~Yang, and S.~Zhu, ``Parallel restarted {SGD} with faster convergence
  and less communication: Demystifying why model averaging works,'' in
  \emph{Proc. AAAI Conference on Artificial Intelligence}, 2019, pp.
  5693--5700.

\bibitem{bagdasaryan2020backdoor}
E.~Bagdasaryan, A.~Veit, Y.~Hua, D.~Estrin, and V.~Shmatikov, ``How to backdoor
  federated learning,'' in \emph{Proc. International Conference on Artificial
  Intelligence and Statistics}, 2020, pp. 2938--2948.

\bibitem{geiping2020inverting}
J.~Geiping, H.~Bauermeister, H.~Dr{\"o}ge, and M.~Moeller, ``Inverting
  gradients-how easy is it to break privacy in federated learning?'' in
  \emph{Proc. Neural Information Processing Systems (NIPS)}, 2020, pp.
  937--947.

\bibitem{wei2021gradient}
W.~Wei and L.~Liu, ``Gradient leakage attack resilient deep learning,''
  \emph{IEEE Trans. Information Forensics and Security}, vol.~17, pp. 303--316,
  2021.

\bibitem{li2020secure}
Y.~Li, T.-H. Chang, and C.-Y. Chi, ``Secure federated averaging algorithm with
  differential privacy,'' in \emph{Proc. IEEE International Workshop on Machine
  Learning for Signal Processing (MLSP)}, 2020, pp. 1--6.

\bibitem{dwork2014algorithmic}
C.~Dwork, A.~Roth \emph{et~al.}, ``The algorithmic foundations of differential
  privacy.'' \emph{Foundations and Trends in Theoretical Computer Science},
  vol.~9, pp. 211--407, 2014.

\bibitem{wang2021clustering}
S.~Wang, T.-H. Chang, Y.~Cui, and J.-S. Pang, ``Clustering by orthogonal {NMF}
  model and non-convex penalty optimization,'' \emph{IEEE Trans. Signal
  Processing}, vol.~69, pp. 5273--5288, 2021.

\bibitem{stallmann2022towards}
M.~Stallmann and A.~Wilbik, ``Towards federated clustering: A federated fuzzy $
  c $-means algorithm {(FFCM)},'' \emph{arXiv preprint arXiv:2201.07316}, 2022.

\bibitem{kolluri2021private}
A.~Kolluri, T.~Baluta, and P.~Saxena, ``Private hierarchical clustering in
  federated networks,'' in \emph{Proc. ACM SIGSAC Conference on Computer and
  Communications Security}, 2021, pp. 2342--2360.

\bibitem{ghosh2019robust}
A.~Ghosh, J.~Hong, D.~Yin, and K.~Ramchandran, ``Robust federated learning in a
  heterogeneous environment,'' \emph{arXiv preprint arXiv:1906.06629}, 2019.

\bibitem{Sattler20}
F.~Sattler, K.-R. M{\"u}ller, and W.~Samek, ``Clustered federated learning:
  Model-agnostic distributed multitask optimization under privacy
  constraints,'' \emph{IEEE Trans. Neural Networks and Learning Systems},
  vol.~32, no.~8, pp. 3710--3722, 2020.

\bibitem{fraboni2021clustered}
Y.~Fraboni, R.~Vidal, L.~Kameni, and M.~Lorenzi, ``Clustered sampling:
  Low-variance and improved representativity for clients selection in federated
  learning,'' in \emph{Proc. International Conference on Machine Learning
  (ICML)}, 2021, pp. 3407--3416.

\bibitem{Dennis21a}
D.~K. Dennis, T.~Li, and V.~Smith, ``Heterogeneity for the win: One-shot
  federated clustering,'' in \emph{Proc. International Conference on Machine
  Learning (ICML)}, 2021, pp. 2611--2620.

\bibitem{ma2022convergence}
J.~Ma, G.~Long, T.~Zhou, J.~Jiang, and C.~Zhang, ``On the convergence of
  clustered federated learning,'' \emph{arXiv preprint arXiv:2202.06187}, 2022.

\bibitem{ding2016k}
H.~Ding, Y.~Liu, L.~Huang, and J.~Li, ``K-means clustering with distributed
  dimensions,'' in \emph{Proc. International Conference on Machine Learning
  (ICML)}, 2016, pp. 1339--1348.

\bibitem{dwork2006our}
C.~Dwork, K.~Kenthapadi, F.~McSherry, I.~Mironov, and M.~Naor, ``Our data,
  ourselves: Privacy via distributed noise generation,'' in \emph{Annual
  International Conference on the Theory and Applications of Cryptographic
  Techniques}, 2006, pp. 486--503.

\bibitem{erlingsson2019amplification}
{\'U}.~Erlingsson, V.~Feldman, I.~Mironov, A.~Raghunathan, K.~Talwar, and
  A.~Thakurta, ``Amplification by shuffling: From local to central differential
  privacy via anonymity,'' in \emph{Proc. of the Thirtieth Annual ACM-SIAM
  Symposium on Discrete Algorithms}, 2019, pp. 2468--2479.

\bibitem{balle2018privacy}
B.~Balle, G.~Barthe, and M.~Gaboardi, ``Privacy amplification by subsampling:
  Tight analyses via couplings and divergences,'' in \emph{Proc. Neural
  Information Processing Systems (NIPS)}, 2018, pp. 6277--6287.

\bibitem{shen2022performance}
X.~Shen, Y.~Liu, and Z.~Zhang, ``Performance-enhanced federated learning with
  differential privacy for internet of things,'' \emph{IEEE Internet of Things
  Journal}, pp. 1--16, 2022.

\bibitem{li2022network}
Y.~Li, S.~Wang, C.-Y. Chi, and T.~Q. Quek, ``Differentially private federated
  learning in edge networks: The perspective of noise reduction,'' \emph{IEEE
  Network}, vol.~36, no.~5, pp. 167--172, 2022.

\bibitem{li2019convergence}
X.~Li, K.~Huang, W.~Yang, S.~Wang, and Z.~Zhang, ``On the convergence of
  {FedAvg} on {non-IID} data,'' in \emph{Proc. International Conference on
  Learning Representations (ICLR)}, 2020, pp. 1--26.

\bibitem{Kmeans_2012}
B.~Bahmani, B.~Moseley, A.~Vattani, R.~Kumar, and S.~Vassilvitskii, ``Scalable
  {$k$-means++},'' in \emph{Proc. VLDB Endowment}, 2012, pp. 622--633.

\bibitem{PKmeans_2014}
T.~Kucukyilmaz, ``Parallel {$k$-means} algorithm for shared memory
  multiprocessors,'' \emph{Journal of Computer and Communications}, vol.~2, pp.
  15--23, 2014.

\bibitem{DBSCAN_1996}
M.~Ester, H.~P. Kriegel, J.~Sander, and X.~Xu, ``A density-based algorithm for
  discovering clusters in large spatial databases with noise,'' in \emph{Proc.
  Knowledge Discovery and Data Mining (KDD)}, 1996, pp. 226--231.

\bibitem{DKCoreset_2013}
M.-F.~F. Balcan, S.~Ehrlich, and Y.~Liang, ``Distributed $k$-means and
  $k$-median clustering on general topologies,'' in \emph{Proc. Neural
  Information Processing Systems (NIPS)}, 2013, pp. 1995--2003.

\bibitem{Dis_kmeans_2016}
H.~Ding, Y.~Liu, L.~Huang, and J.~Li, ``K-means clustering with distributed
  dimensions,'' in \emph{Proc. International Conference on Machine Learning
  (ICML)}, 2016, pp. 1339--1348.

\bibitem{pedrycz2021federated}
W.~Pedrycz, ``Federated {FCM}: Clustering under privacy requirements,''
  \emph{IEEE Trans. Fuzzy Systems}, vol.~30, no.~8, pp. 3384--3388, 2022.

\bibitem{hernandez2021federated}
E.~Hern{\'a}ndez-Pereira, O.~Fontenla-Romero, B.~Guijarro-Berdi{\~n}as, and
  B.~P{\'e}rez-S{\'a}nchez, ``Federated learning approach for spectral
  clustering.'' in \emph{Proc. European Symposium on Artificial Neural
  Networks}, 2021, pp. 423--428.

\bibitem{li2021federated}
C.~Li, G.~Li, and P.~K. Varshney, ``Federated learning with soft clustering,''
  \emph{IEEE Internet of Things Journal}, vol.~9, no.~10, pp. 7773--7782, 2021.

\bibitem{xu2021asynchronous}
C.~Xu, Y.~Qu, Y.~Xiang, and L.~Gao, ``Asynchronous federated learning on
  heterogeneous devices: A survey,'' \emph{arXiv preprint arXiv:2109.04269},
  2021.

\bibitem{wang2022federated}
S.~Wang and T.-H. Chang, ``Federated matrix factorization: Algorithm design and
  application to data clustering,'' \emph{IEEE Trans. Signal Processing},
  vol.~70, pp. 1625--1640, 2022.

\bibitem{chung2022federated}
J.~Chung, K.~Lee, and K.~Ramchandran, ``Federated unsupervised clustering with
  generative models,'' in \emph{Proc. AAAI International Workshop on Trustable,
  Verifiable and Auditable Federated Learning}, 2022, pp. 1--9.

\bibitem{bonawitz2017practical}
K.~Bonawitz, V.~Ivanov, B.~Kreuter, A.~Marcedone, H.~B. McMahan, S.~Patel,
  D.~Ramage, A.~Segal, and K.~Seth, ``Practical secure aggregation for
  privacy-preserving machine learning,'' in \emph{Proc. ACM SIGSAC Conference
  on Computer and Communications Security}, 2017, pp. 1175--1191.

\bibitem{TCGA_CGCD}
R.~Mclendon, A.~Friedman, D.~Bigner \emph{et~al.}, ``Comprehensive genomic
  characterization defines human glioblastoma genes and core pathways,''
  \emph{Nature}, vol. 455, pp. 1061--1068, 2008.

\bibitem{website_MNIST}
\BIBentryALTinterwordspacing
Y.~LeCun, C.~Cortes, and C.~Burges. The {MNIST} database. [Online]. Available:
  \url{http://yann.lecun.com/exdb/mnist.}
\BIBentrySTDinterwordspacing

\bibitem{abadi2016deep}
M.~Abadi, A.~Chu, I.~Goodfellow, H.~B. McMahan, I.~Mironov, K.~Talwar, and
  L.~Zhang, ``Deep learning with differential privacy,'' in \emph{Proc. ACM
  SIGSAC Conference on Computer and Communications Security}, 2016, pp.
  308--318.

\bibitem{yang2016learning}
B.~Yang, X.~Fu, and N.~D. Sidiropoulos, ``Learning from hidden traits: Joint
  factor analysis and latent clustering,'' \emph{IEEE Trans. Signal
  Processing}, vol.~65, pp. 256--269, 2016.

\bibitem{K++_2007}
D.~Arthur and S.~Vassilvitskii, ``K-means++: The advantages of careful
  seeding,'' in \emph{Proc. Symposium on Discrete Algorithms (SODA)}, 2007, pp.
  1027--1035.

\bibitem{FedProx_2018}
T.~Li, A.~K. Sahu, M.~Sanjabi, M.~Zaheer, A.~Talwalkar, and V.~Smith,
  ``Federated optimization in heterogeneous networks,'' in \emph{Proc. Machine
  Learning and Systems}, 2020, pp. 1--12.

\bibitem{Shuai_SNCP_2019}
S.~Wang, T.-H. Chang, Y.~Cui, and J.-S. Pang, ``Clustering by orthogonal
  non-negative matrix factorization: A sequential non-convex penalty
  approach,'' in \emph{Proc. IEEE International Conference on Acoustics,
  Speech, and Signal Processing (ICASSP)}, 2019, pp. 5576--5580.

\bibitem{SNMF_Kim}
H.~Kim and H.~Park, ``Sparse  non-negative matrix factorization via
  alternating non-negative-constrained least squares for microarray data
  analysis,'' \emph{Bioinformatics}, vol. 23, no. 12, pp. 1495--1502, 2007.

\bibitem{Zhang_ONMF_SM2016}
W.~E. Zhang, M.~Tan, Q.~Z. Sheng, L.~Yao, and Q.~Shi, ``Efficient orthogonal
  non-negative matrix factorization over {Stiefel} manifold,'' in \emph{Proc.
  ACM International on Conference on Information and Knowledge Management
  (ICKM)}, 2016, pp. 1743--1752.

\bibitem{yu2005soft}
K.~Yu, S.~Yu, and V.~Tresp, ``Soft clustering on graphs,'' in \emph{Proc.
  Neural Information Processing Systems (NIPS)}, 2005, pp. 1--8.

\bibitem{FL_Beyond_2015}
J.~K\v{o}necn\'{y}, H.~B. McMahan, and D.~Ramage, ``Federated optimization:
  Distributed optimization beyond the datacenter,'' in \emph{Proc. NeuIPS
  Optimization for Machine Learning Workshop}, 2015, pp. 1--5.

\bibitem{FL_Ondevice_2016}
J.~K\v{o}necn\'{y}, H.~B. McMahan, D.~Ramage, and P.~Richtarik, ``Federated
  optimization: Distributed machine learning for on-device intelligence,''
  \emph{arXiv preprint arXiv:1610.02527}, 2016.

\bibitem{SFMF_2019}
D.~Chai, L.~Wang, K.~Chen, and Q.~Yang, ``Secure federated matrix
  factorization,'' \emph{IEEE Intelligent Systems}, vol. 1, no. 1, pp. 1--8,
  2020.

\bibitem{tseng2001convergence}
P.~Tseng, ``Convergence of a block coordinate descent method for
  nondifferentiable minimization,'' \emph{Journal of Optimization Theory and
  Applications}, vol. 109, pp. 475--494, 2001.

\bibitem{zhang2021fedpd}
X.~Zhang, M.~Hong, S.~Dhople, W.~Yin, and Y.~Liu, ``{FedPD}: A federated
  learning framework with adaptivity to {non-IID} data,'' \emph{IEEE Trans.
  Signal Processing}, vol.~69, pp. 6055--6070, 2021.

\bibitem{ChiBook2017}
C.-Y. Chi, W.-C. Li, and C.-H. Lin, \emph{\BIBforeignlanguage{English}{Convex
  Optimization for Signal Processing and Communications: From Fundamentals to
  Applications}}.\hskip 1em plus 0.5em minus 0.4em\relax CRC Press, Boca Raton,
  FL, Feb. 2017.

\bibitem{BOLTE2014PROXIMAL}
J.~Bolte, S.~Sabach, and M.~Teboulle, ``Proximal alternating linearized
  minimization for nonconvex and nonsmooth problems,'' \emph{Mathematical
  Programming}, vol. 146, pp. 459--494, 2014.

\end{thebibliography}

\end{document}